# Building Human Values into Recommender Systems: An Interdisciplinary Synthesis

**Jonathan Stray[1], Alon Halevy[2], Parisa Assar[2], Dylan Hadfield-Menell[3], Craig Boutilier[4], Amar Ashar[5], Lex Beattie[5], Michael Ekstrand[6], Claire Leibowicz[7], Connie Moon Sehat[8], Sara Johansen[9], Lianne Kerlin[10], David Vickrey[11], Spandana Singh[12], Sanne Vrijenhoek[13], Amy Zhang[14], McKane Andrus[7], Natali Helberger[15], Polina Proutskova[10], Tanushree Mitra[16], Nina Vasan[9]**

[1] Center for Human-Compatible Artificial Intelligence, UC Berkeley, [2] Meta AI, [3] Department of Electrical Engineering and Computer Science, MIT, [4] Google Research, [5] Spotify Inc., [6] Computer Science Department, Boise State University, [7] Partnership on AI, [8] Hacks/Hackers, [9] Brainstorm: The Stanford Lab for Mental Health Innovation, Department of Psychiatry and Behavioral Sciences, Stanford University, [10] BBC, [11] Meta, [12] New America's Open Technology Institute, [13] Faculty of Law, Information Law, University of Amsterdam, [14] Allen School of Computer Science & Engineering, University of Washington, [15] Institute for Information Law, University of Amsterdam, [16] Information and Computer Science, University of Washington

## Abstract

Recommender systems are the algorithms which select, filter, and personalize content across many of the world's largest platforms and apps. As such, their positive and negative effects on individuals and on societies have been extensively theorized and studied. Our overarching question is how to ensure that recommender systems enact the values of the individuals and societies that they serve. Addressing this question in a principled fashion requires technical knowledge of recommender design and operation, and also critically depends on insights from diverse fields including social science, ethics, economics, psychology, policy and law. This paper is a multidisciplinary effort to synthesize theory and practice from different perspectives, with the goal of providing a shared language, articulating current design approaches, and identifying open problems. It is not a comprehensive survey of this large space, but a set of highlights identified by our diverse author cohort. We collect a set of values that seem most relevant to recommender systems operating across different domains, then examine them from the perspectives of current industry practice, measurement, product design, and policy approaches. Important open problems include multi-stakeholder processes for defining values and resolving trade-offs, better values-driven measurements, recommender controls that people use, non-behavioral algorithmic feedback, optimization for long-term outcomes, causal inference of recommender effects, academic-industry research collaborations, and interdisciplinary policy-making.

## 1. Introduction

Recommender systems are the algorithms which select, filter, and personalize content across social media (Lada et al., 2021), news aggregators (Das et al., 2007), music and video streaming services (Holtz et al., 2020; Zhao et al., 2019), online shopping (Louca et al., 2019), online ad targeting (Zhao et al., 2021), and other systems. As such, their positive and negative effects on individuals and societies have been extensively theorized and studied. In the context of social media recommendation there has been mixed evidence regarding both positive and negative effects on adolescent well-being (Przybylski and Weinstein, 2017), polarization (Allcott et al., 2020; Asimovic et al., 2021) and news consumption (Fletcher and Nielsen, 2018). In news recommender systems, diversity of opinion and content sourcing is a major concern (Bernstein et al., 2020; Helberger et al., 2018). Recommender systems used to promote job openings may be discriminatory if they do not consider the balance of distribution across legally protected user attributes (Lambrecht and Tucker, 2019). Product recommendations used for online shopping could shift large-scale behavioral patterns with significant economic, environmental, or social effects (Gretzel and Fesenmaier, 2006). Even systems designed purely for entertainment, such as film and music streaming services, must consider the fair allocation of attention to the artists who create content (Mehrotra et al., 2018). One overarching question across all of these contexts is, *how can we make recommender systems enact the values of the individuals and societies that they serve*?

Addressing this question in a principled fashion requires technical knowledge of recommender design and operation, and also critically depends on insights from diverse fields including social science, ethics, economics, psychology, and law. This paper is a multidisciplinary effort to define a common language and review the current state of the art for addressing problems related to designing and operating recommender systems in support of a wide array of human values. This is a wide-ranging exercise in value-sensitive design (Friedman et al., 2002), which we undertake by synthesizing knowledge from a variety of perspectives.

There are several widely used frames for discussing the normative implications of AI systems, most of which apply to our narrower context of recommender systems. "Alignment" is concerned with ensuring that AI systems enact the intentions of its designers and of its users despite the impossibility of specifying the correct action in all possible cases (Hadfield-Menell and Hadfield, 2019) and provides us with the language of "value alignment." "Fairness" or "bias" is primarily concerned with the distribution of benefits and harms between people or groups (Chen et al., 2020; M. D. Ekstrand et al., 2021). "Integrity" refers to identifying and moderating content that violates platform policies for a variety of reasons, including obscenity, copyright, criminal activity, and misinformation (Halevy et al., 2020). "Well-being" is an umbrella term for a wide variety of sociological measures used across sectors including government, public health, and research, which are starting to be applied to AI systems (Schiff et al., 2020). There is a body of work on the "ethics" of recommender systems (Milano et al., 2020). The potential effects of AI systems have also been analyzed within a "human rights" framework (Donahoe and Metzger, 2019). There is substantial overlap between these categories, each of which encompasses a range of more specific concerns such as misinformation (Fernandez and Bellogín, 2020), polarization (Stray, 2021a) or addiction (Allcott et al., 2021).

Rather than trying to reconcile these diverse frameworks, we take all of these to be concerned with *values*. We draw on the field of value-sensitive design (Friedman et al., 2002) to define values as "what a person or group of people consider important in life" (Borning and Muller, 2012, p. 1). Building real systems requires both deep technical practice and grounding in the realities of diverse human lives, including an understanding of the actual effects of deployed recommenders. Further, values and our understanding of them are constantly evolving, so the whole exercise depends on moral and philosophical reflection. Therefore, building human values into recommender systems requires a mix of conceptual, empirical, and technical work (Friedman et al., 2017).

Recommender systems are fundamentally social and collaborative both because many of their applications are designed to facilitate user-to-user interactions (e.g. social media) and because most recommender algorithms work by identifying patterns across users (e.g. "people who liked this also liked that"). Value-sensitive design uses the language of *stakeholders,* defined as those who are affected by design choices. It is their values which should matter in the design process, and recommenders are fundamentally multi-stakeholder systems which must simultaneously serve several types of stakeholders including users, content producers, platforms, and non-users (Abdollahpouri et al., 2020). Their opinions must be considered or solicited in some way, and there are a wide variety of methods for doing so in various contexts including participatory design, user surveys, semi-structured interviews, and deliberative elicitation (Lee et al., 2019; Ovadya, 2021; Simonsen and Robertson, 2012; Stray, 2020). To the extent that this feedback influences recommender algorithms, this is also an example of value-sensitive algorithm design (Zhu et al., 2018)

This paper is not a comprehensive review. We do not think such a review would be useful at this time, both because it would be massive and because the field is evolving rapidly. Rather, the goal of this paper is to help readers get their bearings in this large and complex space. While not exhaustive, our approach is interdisciplinary and cross-sector. Our authors include experts in recommender systems, journalism, political science, law, psychology, AI alignment, technology policy, human computer-interaction and other fields. We also have substantial representation across academic, industry, and civil society sectors. Even this broad author cohort cannot reflect the entire range of this space, so we focus primarily on recommenders for social media, news content, and entertainment streaming services. However, many of the issues and approaches we highlight are broadly applicable to other important categories of recommender systems, including online shopping, targeted advertising, recruitment, healthcare, and education.

We use the term "recommender systems" to focus on the core problem of personalized content selection across many domains. Recommender systems often operate without an explicit user query, though the user may also ask for more tailored recommendations (e.g., "politics podcast"). This contrasts with search functionality which requires an explicit query and where results tend to be much less personalized (Courtois et al., 2018; Krafft et al., 2019; Le et al., 2019). Social media is a major application of recommender systems, but we note that the two are not synonymous. While most social media platforms do employ recommenders for content selection, the effects of social media depend on many design decisions beyond algorithmic content selection including the way the content is displayed, the various possible actions or controls made available to users, and the use of other navigation methods such as search or human-curated lists. Further, it is difficult to disentangle recommender effects from user creation, consumption, and sharing behavior across multiple platforms (Bartal et al., 2020; Hosseinmardi et al., 2020; Munger and Phillips, 2020). Nor do we take up content moderation at any length, even though content removal intersects strongly with values (Klonick, 2018; York and Zuckerman, 2019) and there is technical overlap with recommender algorithms. Instead we are concerned with what content *is* selected after items ineligible for recommendation have been removed.

We have been asked why this work is framed in terms of "values" and not "harms." After all, it might seem more urgent to address a harm such as "contributes to depression" than to improve a benefit such as "keeps users happy." In the sense that harms arise when recommenders do not operate in accordance with our values, the framing of values vs. harms can sometimes be merely a linguistic convention. Yet we believe that focusing solely on harms is overly narrow. One widely used set of core principles in applied ethics, deriving originally from consideration of biomedical research in the 1970s, is respect for autonomy, justice, and both avoiding harm and doing good (Beauchamp and Childress, 2019). Considering only harms and not the potential for doing good is limiting, while framing the effects of recommenders in terms of values opens up new avenues for thinking about the role and responsibilities of recommenders in society. Most of the values we consider can span the spectrum from mitigating harm to promoting good.

Further, it may not be possible to build a system that avoids "bad" without also defining "good," because of the necessity of specifying objectives. A news recommender should not promote false information, but this does not provide any guidance on what sort of information it *should* select, nor how it should support the societal goals of journalism (Fields et al., 2018; Helberger, 2019). Similarly, if social media recommenders should not optimize for engagement, then what should they optimize for? This question is more than rhetorical, as most AI technologies are based on optimizing some set of metrics which define the objectives of the system, and the appropriate specification of objectives is a significant and well-known challenge (D'Amour et al., 2020; Gabriel and Ghazavi, 2021; Hadfield-Menell and Hadfield, 2019; Mullainathan and Obermeyer, 2021).

We do not attempt to offer a general theory of recommender value alignment. Instead, our approach is to synthesize current thinking, with an emphasis on industrial applicability. The contributions of this paper include:

- A set of human values that are important in the context of recommender systems, derived from previous partial compilations and feedback from experts in ethics, policy, and industry. We summarize the state of research and practice around each of these values, including the empirical effects of existing recommender systems (Section 2).

- We describe one possible process for operationalizing a given human value, based on current practice in industrial recommender systems, through the example of increasing the political diversity of content in a production news recommender (section 3).

- We discuss the techniques used to measure adherence and progress on human values. In particular, we show how measurement techniques of social science apply to the problem of evaluating values-based outcomes in recommender systems (section 4).

- We describe known design approaches to altering recommender systems to conform to specific values, which provides a menu of possibilities to recommender designers and opportunities for future work (section 5).

- We consider the different approaches to policy-making related to human values, and identify gaps where the policy and technical communities need to align (section 6).

- Finally, we present a set of important open problems across all the above areas (section 7).

While a work of this scope cannot possibly be comprehensive, this is a synthesis based on our collective understanding and experience, the result of an extensive interdisciplinary conversation. We offer a view of the current state of thinking and practice across a wide variety of issues and applications, and suggest the most promising directions for future work.

## 2. Values for Recommender Systems

In this section we consider a variety of values applicable to recommender systems. We do not aim to create a comprehensive list of all relevant values (which may not even be possible!). Instead, our goal is to ground the many discussions happening in industry, academia, and policy in a reasonably broad set of values so progress can be made. We used a multi-source methodology to come up with this list, incorporating previous reviews of values in technology, expert deliberation, and a small qualitative expert survey.

We began with four substantial compilations of values and risks relevant to either recommenders in particular or AI systems in general. These were a survey of AI ethics policy documents (Fjeld et al., 2020), a large user research project undertaken by the BBC (Kerlin, 2020), a compilation of AI-relevant well-being metrics (IEEE, 2020), and an IEEE process for ethical system design (IEEE, 2021). We extracted and combined these lists, then added a set of global "techno-moral virtues" (Vallor, 2016, pp. 15,120) as well as the feminist ethic of care (Skoe, 2014) and philosophical traditions in Africa that emphasize the relationships and bonds among people (Ndiweni and Sibanda, 2020). We added a few more items by hand, according to our judgment as practitioners and researchers, to ensure our list includes known "issues" in recommender operation (e.g. polarization).

In order to refine our list and better understand the trade-offs between different values, we gathered multi-stakeholder input through the Partnership on AI (PAI), a global non-profit partnership of roughly 100 academic, civil society, industry, and media organizations working towards positive outcomes from AI for people and society. PAI surveyed those of its members with an interest in recommender systems, approximately 40 people, with this initial list of values asking them first to choose a type of recommender system (e.g., social media, news aggregator) then to rate the importance of each value for that recommender system. At a subsequent workshop, these 40 participants discussed these values in small groups to further develop their conception of how values apply in recommender systems (Leibowicz et al., 2021). We then refined our list and definitions of values based on this feedback.

The goal of this exercise was not to arrive at a set of global, universally applicable values, but to list some of the key values that have been identified to be particularly relevant in various recommender contexts. We note that the main challenge in devising our list was coming up with a set of definitions at an appropriate level of granularity that together cover a wide set of overlapping and often vague concepts. We aimed for a level of abstraction between overly general statements (e.g., "do good") and specific formulations (e.g., particular metrics). Our list is shown in Appendix A, along with example indicators or metrics for each value, and example design changes which could promote that value.

Values cannot be discussed in isolation. By nature, there are tensions between values, and those tensions lead to many of the difficulties in operationalizing them in recommender systems. For example, the value of *free expression* is in tension with the value of *safety* because if we allow users to say anything they want on social media, others may feel threatened. As another example, *privacy* can be in tension with *usefulness*. *Privacy* suggests that a platform should not try to infer whether a user might have a particular disease, even though early information and intervention might be helpful to them.

Some of these tensions are rooted in the tradeoff between individual values versus societal values. For example, if we exclusively focus on the *well-being* of individuals, society may suffer (Ostrom, 2000). Values can also have effects on varying time scales. Giving users entertaining content may satisfy their short term needs, but providing more informative content may have longer term benefits. Having informed users may also be a societal value, and individual values often operate on a shorter timescale than societal values.

Many of the values in our list are closer in nature to instrumental goals for an AI system, by which higher-order values are achieved. For example, the bioethics principles of respect for *autonomy*, *beneficence*, and *justice* are often considered primary, without resorting to other values for justification (Canca, 2020). By contrast, other values seem to derive much of their importance because they contribute to these principles, including *privacy, agency, control, transparency, accessibility/inclusiveness,* and *accountability*.

Rather than discussing each value individually, we approach them through a number of themes that characterize many discussions on these topics: usefulness, well-being, legal and human rights, public discourse and safety.

## Usefulness

Platforms use recommender systems because they believe them to be useful to users, content creators, and themselves. The most straightforward distinction between recommendation and search is that a recommender can suggest items without an explicit query, which is valuable in a variety of contexts. For example, news items cannot be selected through user queries alone, because the user is unaware of new events. While the value of presenting a previously unknown post, article, person, movie, song or ad varies, all of these can lead to positive and novel outcomes for people. Modern recommender systems have their roots in collaborative filtering systems in the 1990s, and the need for intelligent filtering has only increased since then as the pool of available information has exploded. We call this value "*usefulness*" to distinguish it from "utility" which has a more technical meaning from economics (discussed section 5.3.1).

*Usefulness* is closely related to *control*, which we take to mean that users should be able to select which content they are seeing, the type and degree of personalization (if any), and understand the processes that determine what they see. In large part this requires that platforms provide features to support such choices (e.g., playlists on a music streaming service or topic selection on a news service) though there are also important questions about community governance (Lee et al., 2019; Zhang et al., 2020). On social networks, interfaces for describing which posts to see are particularly complex given the breadth of available content. Note that *feeling in control* and *actually having control* are different. Placebo controls may increase user satisfaction without offering any actual improvement (Vaccaro et al., 2018) while users may not perceive even large effects of functional controls (Loecherbach et al., 2021). *Agency* is a similar value, but we use it to refer to control over other elements of the users' life, not just the recommender. For example, a recommender could assist a user with education (Dascalu et al., 2016) or direct them to job opportunities.

There are complex tensions between *agency* and *control* and other values, as users might make choices that harm themselves or others. This grounds out in concrete ethical questions such as: if it is possible to infer that someone has an eating disorder (Yan et al., 2019), and if there is research indicating that viewing dieting videos leads to bad outcomes for such a person, is it reasonable or even obligatory to thwart their expressed intention to see such material?

Recommender systems can also be useful to both content creators and platforms. While recommenders are a commercially important technology, algorithmic content optimization does not translate directly into revenue in many contexts (Jannach and Jugovac, 2019). For example, subscription services must maximize user retention, while current recommender designs struggle with long-term outcomes. Direct optimization for revenue, or more precisely profit, is best developed in the context of online shopping (Das et al., 2009; Delgado et al., 2019; Jannach and Adomavicius, 2017; Louca et al., 2019). Even for ad-driven services, content personalization and ad selection are often handled by different recommenders which optimize against different objectives (Thompson, 2018). Other organizations which operate recommenders may not intend to make money by doing so, such as news publishers. While not all recommender systems need to be profitable, they all must be useful to the operator by some measure.

### Well-being

*Well-being* is fundamental to human experience, and recommender systems have the potential to affect users' *well-being* in many ways. While the phrase has specific implications around positive subjective experience, it is also widely used in the policy community as an umbrella framework which encompasses other values (Exton and Shinwell, 2018; Graham et al., 2018; O'Donnell et al., 2014) and has been explored as an important end goal for AI systems in general (IEEE, 2020). *Well-being* as a subjective experience is one of the values in our list, but many other values intersect with it strongly and are often included in well-being frameworks.

Well-being is a complex concept, and there is little consensus on how to define it (Dodge et al., 2012). Objective measures such as employment, lack of crime, and economic prosperity were historically used as proxies for *well-being*. More recently there has been increasing focus on a more holistic understanding of *well-being* based on both subjective and objective measures (Diener, 2000; Hicks et al., 2013; Krueger and Stone, 2014). These subjective *well-being* measures account for people's cognitive and affective evaluations of their lives by asking subjects to rate how much they agree with statements like, "The conditions of my life are excellent" (Diener et al., 1985).

The values of *connection*, *community and belonging*, *recognition and acknowledgement*, *self expression*, *care*, *compassion*, and *empathy* all relate to the concept of *well-being*. Many different types of content can contribute to increased *well-being*, such as through education, motivation, or personal relationships. *Entertainment* can also contribute to *well-being*, especially as a short-term emotional experience, and hedonism is often considered a basic drive (Schwartz, 2012).

The effect of recommender systems on *well-being* has been most widely studied in the context of social media use. There is conflicting evidence for both positive and negative effects, in part because *well-being* in this context is variously defined and often represented by other factors, from self-esteem and life satisfaction to *civic engagement*, social capital, and user satisfaction. Also, most studies have not been designed to separate the effects of algorithmic content selection from other aspects of social media such as user creation and sharing.

In one study of over 2,000 college students, social media use was associated with improvement in various facets of *psychological well-being* such as overall life satisfaction, *civic engagement*, and social trust (Valenzuela et al., 2009). Some investigators have suggested that the amount of time spent online is less important than the quality of that time; active use may promote well-being, whereas passive use and emotional connection to use may have a negative impact (Bekalu et al., 2019; Frison and Eggermont, 2020; Verduyn et al., 2017). The literature also points toward negative health effects related to social media use. One longitudinal study compared social media use with *mental health*, *physical health*, and found a decrease of 5%-8% of a standard deviation in self-reported *mental health* (Shakya and Christakis, 2017). In another study, deactivating social media accounts for four weeks resulted in increased time in offline interactions and improved subjective *well-being* (Allcott et al., 2020). Upward social comparison has been proposed as one potential link between social media use and *mental health* disorders such as depression and anxiety (Midgley et al., 2021; White et al., 2006).

Some recommender-based products may encourage addictive tendencies. Allcott et al. (2021) found that abstaining from social media use for a time or allowing people to set future screen-time limits produced a decrease in subsequent use, suggesting that social media use may result in habit formation and self-control issues. These effects may not be due to personalized recommendations *per se*, as broadcast television, a non-personalized medium, has also been found to be addictive in this sense (Frey et al., 2007).

There is classic work on the advantages of both strong and weak social ties (Aral, 2016). These benefits are less well studied in the context of recommender systems, but a few studies of social media are worth noting. Social media use has been associated with increased *social connection* and social capital in online and offline social networks, with particular benefit for users experiencing low self-esteem and low life-satisfaction (Ellison et al., 2007; Steinfield et al., 2008; Subrahmanyam et al., 2008). Larger social networks may be associated with greater perceived social support, reduced stress, and improved *well-being* (Nabi et al., 2013). Social media use has been shown to increase intergroup contact and reduce prejudice when offline social-network diversity is low (Asimovic et al., 2021). Recommenders can also contribute to *civic engagement*. A large-scale experiment showed that messaging on social media can increase voter turnout on the order of 1% (Bond et al., 2012). While the possibility that recommendations connect people to violent or extremist groups has been widely discussed (and we review the evidence below) the converse of this concern is the possibility of connecting people to constructive communities or social movements.

We see long term *well-being* as a major open problem for recommender systems, with a number of challenging sub-problems: defining *well-being* in contextually appropriate ways, measuring user *well-being* in production, and algorithmically optimizing for relatively long-term outcomes (months to years).

## Legal and Human Rights

Some of the values in our list can be considered rights, in the sense of obligations to various parties. Previous work has explored the legal and human rights potentially impacted by recommender systems (Milano et al., 2020) and by AI in general (Donahoe and Metzger, 2019)

*Privacy* has at least three different interpretations in the recommender context. The first sense refers to information the system knows, infers or estimates about the user. The second refers to the possibility of direct revelation of user information to other users or third-parties. There is also the possibility of indirect (often noisy) revelation of user information by the actions of the recommender, e.g. a recommendation made to one user may allow them to infer something about the items viewed by another user. This form of information revelation is the subject of a large body of research in the area of differential privacy (Chien et al., 2021; Dwork and Roth, 2014; Jain et al., 2018). *Privacy* trades off against other values such as *usefulness, transparency,* and *fairness.* The question of demographic inference which might help the user is well discussed in the context of algorithmic fairness (Andrus et al., 2021) but analogous concerns arise with many types of user knowledge. Granting users *control* over whether and how they want platforms inferring and using their personal data to recommend useful content might be one way to reconcile *privacy* and ethical use of personal data.

Like *privacy*, *transparency and explainability* are broad concepts. *Explainability* might mean that a recommender system gives reasons why a certain piece of content is being shown to a user (Zhang and Chen, 2020). *Transparency* could require platform disclosure of various types of data including metrics characterizing different types of content or aggregated user activity. *Transparency* can be important for building users' trust (Cramer et al., 2008) and is often discussed as a policy tool to promote understanding and *accountability*. Explanations and disclosures can come in multiple forms depending on their intended audience, such as users, researchers, governance bodies or auditors, and meaningful transparency needs to be informed by concrete individual and societal information needs (Singh, 2020a; van Drunen et al., 2019). From a technical perspective, explanations can be tricky to generate because many recommender techniques (including deep learning) do not lend themselves to easy explanations (Weld and Bansal, 2019).

Fairness, equity and equality are closely tied to human rights (Universal Declaration of Human Rights, UN). In everyday situations where recommender systems are employed, multiple parties have some interest in the outcome (Abdollahpouri and Burke, 2019). Fairness has a variety of meanings in recommender systems, including considering the disparate impacts of recommendations across user classes (Leonhardt et al., 2018). Content providers want to be fairly treated in terms of the exposure and benefit they receive from the system (Diaz et al., 2020; Jeunen and Goethals, 2021); users want to receive good quality of service, and may not want to be under-served relative to other users (Ekstrand et al., 2018); and other stakeholders, such as the system operator, content creators, and society broadly each have ideas of what it may mean for recommendation to be “fair” (Diaz et al., 2020; M. Ekstrand et al., 2021). This has deep ties to various conceptions of *equity*, in particular equity of attention (Biega et al., 2018, p. 201; Singh and Joachims, 2018). Recommenders may also create externalities that affect people who do not use the platform, such as by directing many new people to a formerly obscure place, or encouraging the consumption of products with environmental consequences (Rolnick et al., 2019). The field of multi-stakeholder recommendation has emerged to tackle these dynamics (Abdollahpouri et al., 2020).

Given all the parties involved through the platform or via externalities, *fairness* is a multifaceted problem where different stakeholders have different objectives and needs from the system. Often the desires of different stakeholders are in conflict; not everyone can have exactly what they want, though it might still be possible to give something worthwhile to everyone.

### Public Discourse

Some of the most intense recent discussion around recommender systems has centered around how they affect public discourse. *Accuracy* of information and *diversity* of content are two prime examples.

Access to accurate and factual information supports human decision making and understanding across myriad domains, ranging from healthcare to politics to economics (Hochschild and Einstein, 2015; Seymour et al., 2015). While falsehoods and misleading content have threatened truth for centuries (Arendt, 1972; O’Neil and Jensen, 2020), online user-generated media may have amplified the threats to sensemaking and decision making. For example, some evidence suggests fabricated news articles (as identified by third-party fact checkers) spread significantly faster through Twitter than genuine news articles, especially articles about politics (Vosoughi et al., 2018). It is not clear to what extent this is an effect of recommender algorithms, as opposed to non-algorithmic social media creation and sharing dynamics (Hosseinmardi et al., 2020; Munger and Phillips, 2020). Similarly, COVID-19 misinfo has been common on social media globally, especially early in the pandemic (Al-Zaman, 2021; Brennen et al., 2020) and there is evidence that suggests misinformation has contributed to the spread of the virus via reduced vaccination rates (Loomba et al., 2021; Pinna et al., 2021). This suggests that personalized recommendations may have had a causal effect on disease spread, though this has not been directly studied.

Many authors have argued that designers of news recommenders have editorial responsibility similar to news editors (Fields et al., 2018; Helberger, 2019; Nechushtai and Lewis, 2019; Sørensen and Hutchinson, 2018). Getting the right information to the right people at the right time is a key normative concern that goes well beyond ensuring accuracy, a value that we call "informativeness." We can look to the tradition of public-service journalism to inform recommender design (Fields et al., 2018; Sørensen and Hutchinson, 2018) but personalized news delivery is a new technology and specific editorial theories are still developing. One approach would be to try to deliver a news item if the user previously expressed an interest in a particular topic, if it reports on events that affect their life, or if there is an opportunity for the user to help others (Stray, 2012). These sorts of ideas have yet to be effectively translated into algorithmic terms.

*Diversity* is a value that can be relevant to consumers, content creators, and society in general. In industrial settings diversity has mostly been studied because increased *diversity* typically results in greater user satisfaction and user engagement, at least up to a point (Holtz et al., 2020; Kunaver and Požrl, 2017; Mehrotra et al., 2018; Wilhelm et al., 2018). Many recommenders use diversification algorithms for the practical task of ensuring that users are not continuously shown the same type of item in their recommendations, often implemented as a re-ranking pass (Google, 2020; Ziegler et al., 2005). Other parties also have an interest in diversity. A streaming music or television platform needs to ensure that the long tail of less popular artists or producers have enough exposure to make it worthwhile for them to stay on the platform (Mehrotra et al., 2018), and increased diversity may contribute to equity of attention (Biega et al., 2018; Singh and Joachims, 2018).

The experience of a lack of diversity in personalized content consumption has been described as a "filter bubble" or "echo chamber." However, this language is somewhat vague and has been used to describe a wide variety of phenomena including self-selected consumption behavior, homophily in social networks, and algorithmic feedback effects (Bruns, 2019). We consider such possibilities more specifically throughout the rest of this paper.

*Diversity* has been most specifically studied in the context of news recommendations, where it might serve a variety of democratic goals including *consumer choice, civic participation, pluralist tolerance*, or challenging the status quo (Bernstein et al., 2020; Helberger, 2019; Vrijenhoek et al., 2020). The meaning of diversity can differ between news organizations, depending on their editorial missions and the balancing of other values that matter to the organization (such as personal relevance, engagement, or time spent). In practice diversity is often measured using item-similarity metrics (Kunaver and Požrl, 2017) but such formulations do not yet capture the complexity that the social sciences have brought to the debate about media diversity (Loecherbach et al., 2020).

One concern is that a lack of diversity in personalized news recommendations could prevent users from being exposed to contrary perspectives. Simulations have suggested that optimization to increase user engagement could create feedback loops that drive users into narrower selections of content (Carroll et al., 2021; Jiang et al., 2019; Kalimeris et al., 2021; Krueger et al., 2020; Rychwalska and Roszczyńska-Kurasińska, 2018). However, news personalization algorithms do not seem to produce a less diverse selection than human editors (Möller et al., 2018) and the news provided to different users is quite similar (Guess et al., 2018; Nechushtai and Lewis, 2019). Social media users consume a more diverse range of news sources than non-users (Fletcher and Nielsen, 2018) but correspondingly also consume more news from partisan sources (Fletcher et al., 2021).

A related concern is that recommender systems might be causing large-scale polarization of ideology (issue polarization) or attitudes (affective polarization) (Stray, 2021a). The evidence for this hypothesis is mixed. In the U.S., polarization began increasing decades before social media (Boxell et al., 2017) while several other developed countries have similar internet usage but do not show increasing polarization (Boxell et al., 2020). Paying people to stop using social media for several weeks produced small declines in issue polarization measures in a study of American users (Allcott et al., 2020). However, an attempted replication in Bosnia and Herzegovina showed *increases* in a measure of ethnic affective polarization after not using social media for two weeks (Asimovic et al., 2021). This was mediated by the ethnic diversity of users' offline social networks, suggesting that in this context the platform was providing otherwise absent exposure diversity.

Finally, we note that there are additional values on our list that have more of a societal flavor such as *civic engagement, social progress, meaningful work, environmental sustainability*, and *tradition and history*. These are important values, though the degree to which they are considered important varies considerably from one culture to another. In our survey and the literature we reviewed these values did not come up as particularly relevant to online platforms, but that does not mean that they cannot or should not be promoted in certain contexts.

## Safety

*Safety* includes the idea that people should not be bullied, attacked, or dehumanized and should not be exposed to disturbing content. For social media, safety has thus far mostly been considered in the context of content moderation, and many platforms have developed their content moderation policies based on international human rights principles and frameworks and in consultation with third party experts (Douek, 2020; Facebook, 2021a; Patja Howell, 2021). As a notable example, a series of human rights abuses occurred in Myanmar, characterized by hate speech and disinformation against the minority Rohingya by the military on social media (BSR, 2018).

Recommender systems should not promote violence. Note that "polarization," a mass hardening of political divisions, is conceptually distinct from "radicalization," where a small number of individuals violate mainstream norms and may resort to violence (van Stekelenburg, 2014). There are a number of documented cases of far-right and Islamist radicalization where online recommendations were involved (Baugut and Neumann, 2020; Munn, 2019; Roose, 2019; Washingtonian, 2019). However, these reports mention many other factors including chat rooms, personal relationships, user-directed searches, and life circumstances. More systematic studies have looked for recommender feedback effects that move users toward radicalization en mass (Faddoul et al., 2020; Ledwich and Zaitsev, 2019; Munger and Phillips, 2020; Ribeiro et al., 2020a) These studies generally show that recommenders alter content mix in the direction of engagement, but have produced poor evidence on the radicalizing potential of recommender systems because of insufficiently powerful experimental designs, as we will discuss below. In general, causal understanding of user trajectories through recommender systems remains a major challenge.

# 3. The State of Practice

To ground the discussion of human values in recommender systems, we review the current state of practice from two perspectives. In Section 3.1 we describe a general architecture that captures common design patterns in current commercial recommender systems. In Section 3.2 we illustrate how the engineering of a recommender system to support a specific value might proceed, within the context of a large company. We note that recommender objectives are chosen and evaluated by people working in teams, which makes these human processes an important determinant of system outcomes. These sections draw heavily on the experience of those of our authors who are practitioners working on large, contemporary recommender systems.

## 3.1 How Recommenders Work

Recommender systems are often described as “black boxes” (Singh and Doty, 2021a) but most are constructed using similar principles. This section presents a greatly simplified, but illustrative recommender design. While it doesn’t represent the details of any particular system, many real systems share its features. Our discussion also leads to a technical definition of *engagement* in terms of behavioral data. Most recommenders are built to optimize some form of engagement at their core, though they also consider many other types of signals.

Recommendations start with a pool of content items. These items may be produced entirely by the recommender operator, as with a news organization’s recommendations; or they may be curated from multiple sources, such as a music recommender which gets content from publishers; or items may be entirely user generated and posted without prior review, as on social media. These three categories have been called “closed,” “curated,” and “open” recommender systems (Cobbe and Singh, 2019). Before recommendations are generated, a moderation process identifies and removes items which violate platform policies. Platform moderation is a complex process which involves many human and machine steps from policy making to enforcement to appeals (Halevy et al., 2020; York and Zuckerman, 2019) but here we are concerned only with how moderation defines which items are available for recommendation.

Individual recommendations are generated using data about item content, user attributes, and context, and result in an output stream or set of recommended items, sometimes called a *feed* or *slate*. The context may include a wide variety of features such as the video the user is currently watching, the time of day, or a search query like “politics podcast” (Bauer and Novotny, 2017; Beutel et al., 2018). User attributes may be derived from personal information the user has provided, any explicit user feedback or control settings, and any implicit feedback contained in the history of past interactions with the system.

The recommendation process usually begins with the selection of *candidate items*. Candidate generation algorithms are tuned to retrieve an overbroad sample, but are very efficient at filtering a corpus of (potentially billions of) items available for recommendation down to a small set which might be a good fit for the user and context, typically ranging in size from a few hundred to a few thousand. These candidates are then *ranked,* that is, each is assigned a *relevance score,* which typically reflects a prediction of user engagement with the candidate item. In modern systems ranking may involve dozens or hundreds of "signals" which summarize aspects of the content, the user, the context, and how all of these interact. The top scoring items are then selected as the user's recommendations. Many systems then re-rank the remaining items, this time comparing them to each other, rather than evaluating each individually, to achieve goals such as diversity of item topic or source (Goodrow, 2021; Ziegler et al., 2005).

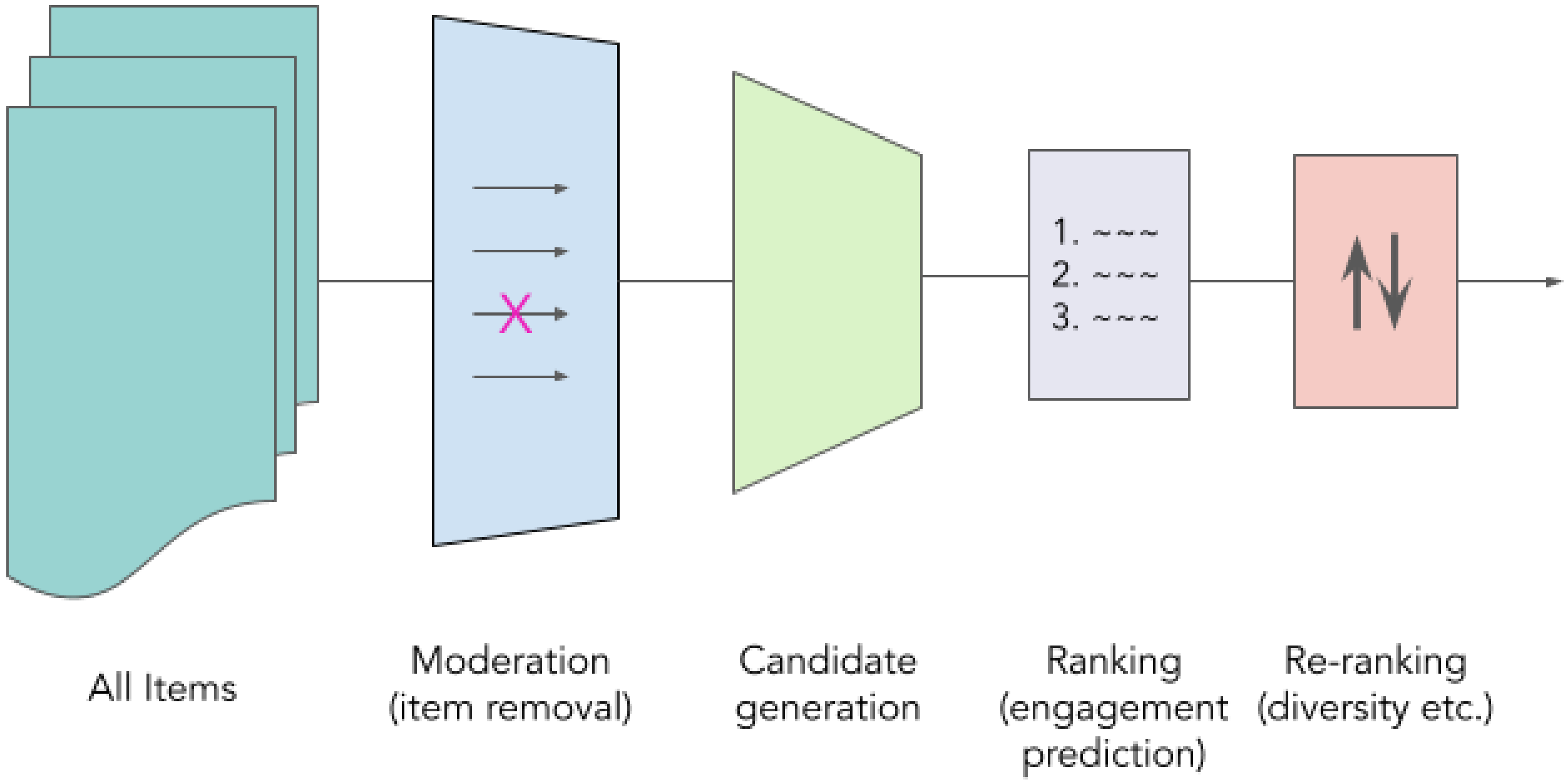

Figure 1: An illustration of how many modern recommenders work, adapted from (Google, 2020; Lada et al., 2021). The global set of items is first moderated to remove content that violates platform policies. The remaining steps happen when a user is served recommendations: candidate generation selects a wide set of items that could be relevant, ranking scores each one, and re-ranking ensures feed-level properties like diversity.

Item ranking constitutes the core of personalization. The final item score is typically a weighted combination or some other function of (i) the predicted probability of a number of different types of user responses (Facebook, 2021b; Milli et al., 2020; Smith, 2021; Zhao et al., 2019) plus (ii) a wide variety of scoring signals that range from source credibility (Pennycook and Rand, 2019) to playlist diversity (Mehrotra et al., 2018) to whether an item tends to be inspiring (Ignat et al., 2021). Ranking items by the probability of desired or targeted user reactions (e.g., sharing, dwell time) is informally known as optimizing for *engagement*. This may or may not optimize for *value* to various stakeholders, which is why many non-engagement signals are also used.

The word "engagement" has been used across many fields including media and technology to suggest that users are repeatedly interacting with a product, as evidenced by a wide variety of metrics. Here we propose a more specific definition, compatible with recommender design practice. We take engagement to be *a set of user behaviors, generated in the normal course of interaction with the platform, which are* *thought to correlate with value to the user, the platform, or other stakeholders*. This definition builds on previous work (Jannach and Adomavicius, 2016; Yi et al., 2014), is multi-stakeholder in nature, and reflects the fact that engagement signals are chosen to be indicators of value, but aren't going to be fully aligned with specific values in all cases. It also suggests there are some signals of value that can only be derived from non-ordinary or non-behavioral data, as we discuss below.

### 3.2 Implementing a Product Change

In this section, we illustrate the process by which a specific human value might be incorporated into a recommender system's design using the value of *diversity*. We walk through the steps that could be involved when a news recommender operator wants to increase political diversity, and illustrate the complex interplay between social, technical and organizational factors when building real products to reflect specific values. It introduces basic themes that we discuss throughout the rest of this paper, including selecting values, creating metrics, implementing product changes, and evaluating trade-offs.

While we hope this example is illuminating, it is not intended to be a general theory of value-sensitive design for recommender systems. Rather, this section is intended to be descriptive, based on the authors' collective experience working with platform-scale recommenders. Unfortunately, there is little public documentation of the actual processes by which values are engineered into large recommender systems. The processes outlined in this example do not represent those of any particular company or product, nor are they universal; but the example nonetheless offers a window into typical practice.

Consider the development or refinement of a news recommendation platform in which the designers are concerned with users having the opportunity to develop an awareness and understanding of multiple political perspectives—this has been called the *deliberative perspective* on recommender diversity (Helberger et al., 2018). The question of which values matter most for any given recommender is an important one, but we suppose the decision has been made to prioritize this type of civic tolerance, perhaps from the study of existing recommender outcomes. One way that a recommender system might contribute to this value is by increasing the diversity of recommended items. This strategy rests on two key assumptions: there is currently a lack of diversity in the items users see, and showing them more diverse items will increase tolerance. Both of these assumptions are complex and available evidence is mixed (Stray, 2021a). Nonetheless, this is a non-trivial example of how metric-driven recommender engineering might proceed.

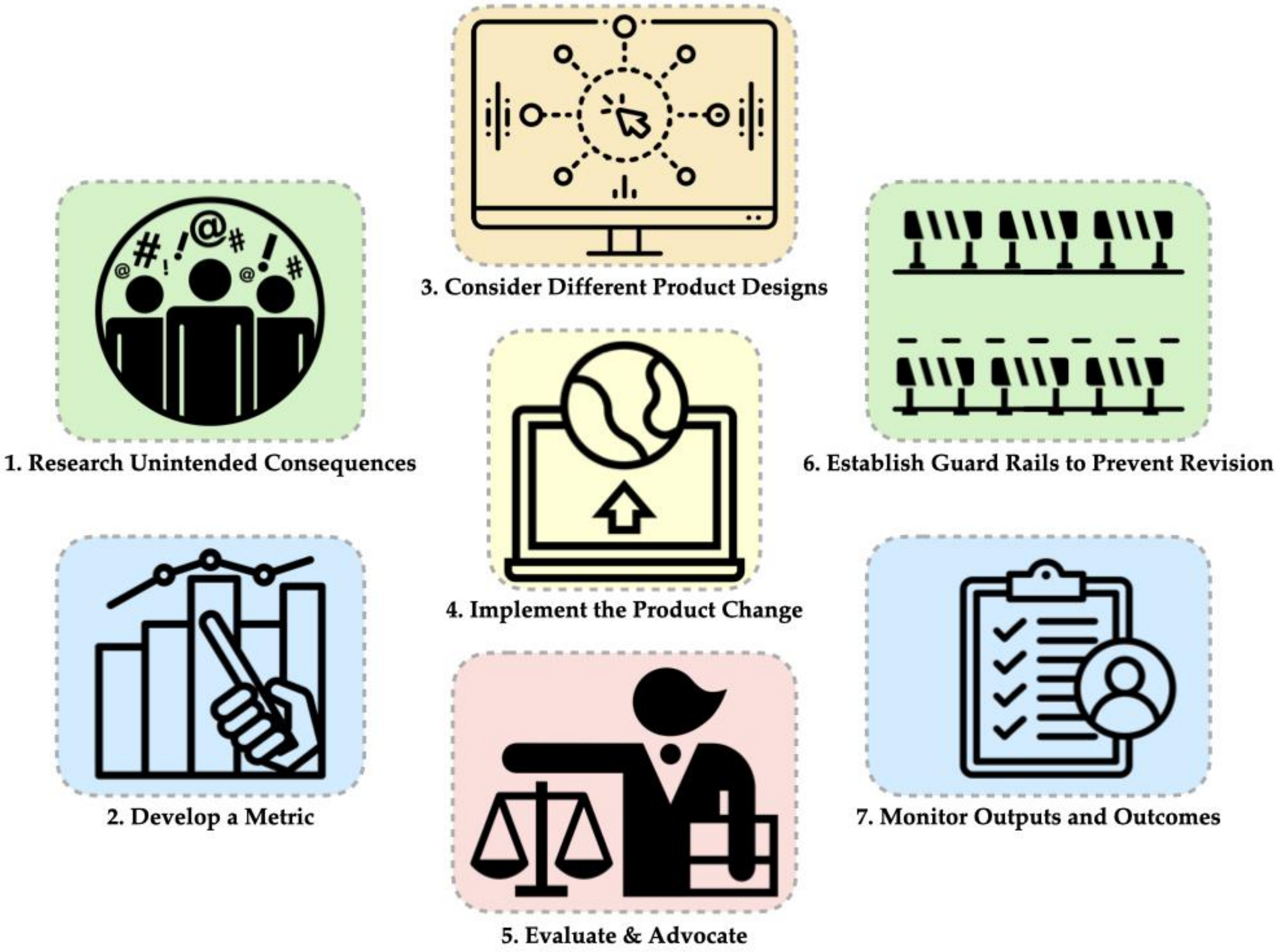

Figure 2: To ground our discussion of values in recommendation systems, this section discusses a potential product change to improve the diversity of content recommendations. We present seven distinct steps: 1) research unintended consequences, to reduce the risk associated with the change; 2) develop a metric to measure progress; 3) consider different product designs; 4) implement the product change; 5) evaluate the change and advocate for it with internal stakeholders; 6) establish guard rails to prevent revision based on other changes; and 7) monitor outputs and outcomes to identify, e.g., drift between your metrics and the concept you ultimately care about.

Once the decision to increase content diversity has been made, the implementation might proceed using the steps outlined below.

#### 3.2.1 Research unintended consequences

Generally, such an effort begins with the study of potential mechanisms to incorporate this value. In the case of diversity, designers may research the history of attempts to create inter-group tolerance through diversity of exposure, including the ways in which this has failed (Paolini et al., 2010; Pettigrew and Tropp, 2006). User studies may be used to test the exposure diversity hypothesis with users of the platform, assessing the impact of seeing news items from diverse perspectives on people's knowledge and attitudes. In particular, it will be important to test for backfire effects where people reject diverse information (Bail et al., 2018; Dunn and Singh, 2014) or other unintended consequences. There must also be a reasonable expectation that the changes to the recommendation system won't adversely affect other values that the system embodies. In concrete terms, this often means that any proposed change should not decrease other metrics (see below) more than a specified amount.

#### 3.2.2 Develop a metric

As discussed in Section 4, metrics are central to any attempt to build values into recommender systems. A variety of diversity metrics have been proposed, both for news content (Cowlishaw et al., 2018; Vrijenhoek et al., 2020) and for recommender systems more generally (Castells et al., 2015; Kunaver and Požrl, 2017). Choosing or developing a metric requires settling on a definition of diversity suitable for that particular system. It may be that multiple metrics are needed to capture different aspects of diversity. While product teams typically choose metrics, there has been experimentation involving external stakeholders in order to increase the legitimacy of such decisions (Bernstein et al., 2020; Lee et al., 2019; Stray, 2020).

Here we focus on a conception of *"productive" diversity*, where people disagree in ways that are ultimately constructive (Stray, 2021a). Given this concept, developing a method to measure the diversity of articles on the platform may be broken into several phases, such as:

1) Developing a description of diversity for use by human raters (e.g., "Does this set of articles include constructive contributions from multiple perspectives?").
2) Creating a training data set of positive and negative examples using human-rater evaluations.
3) Using this training data to develop a heuristic or machine-learned model that can predict whether a list of (recommended) articles adequately reflects "productive diversity."

#### 3.2.3 Consider different product designs

Before implementing a specific product change, different product-based approaches to increasing diversity will be explored. For example, the designer could change the user interface, perhaps by showing related articles from other viewpoints below each item (Su, 2017), or change the ranking algorithm to try to nudge people to consume more articles that represent this type of diversity (Mattis et al., 2021). For the sake of example, we assume below that the latter change has been selected.

#### 3.2.4 Implement the product change

Implementing this specific product refinement requires incorporating the diversity prediction model into the item ranking procedure. Whereas many current recommender systems score each item independently, diversity is a property of sets of items. One challenge is that scoring entire lists is both more complicated and more expensive than scoring individual items (Ie et al., 2019; Wilhelm et al., 2018) which may necessitate the development of more efficient algorithms for ranking such item sets.

#### 3.2.5 Evaluate and advocate

Once implemented, the new product feature will typically be tested using offline data (Cañamares et al., 2020; Gilotte et al., 2018) followed by A/B tests with a small groups of users. If these tests show positive results, the new diversity prediction model will be deployed in production, and monitored to see whether the target diversity metric improves (this may involve gradual ramping up of the deployment to larger numbers of users, holdback tests, etc.). There may also be side effects—for example, while increasing diversity often increases engagement (Kunaver and Požrl, 2017; Wilhelm et al., 2018), this is not always the case. Suppose that as a by-product of the model deployment there is a drop in engagement, declines in other values-relevant outcomes such as quality or safety metrics, or some cost imposed on other stakeholders (e.g., content producers). In practice, this requires negotiation among internal stakeholders to decide if the increase in diversity is significant enough to justify a drop in other metrics.

#### 3.2.6 Establish guardrails to prevent reversion

Once deployed, product teams often establish a review process so that subsequent product changes, within the originating team or elsewhere, don't indirectly revert the diversity improvement. This might include numeric "guardrails" that specify the maximum allowable decrease in diversity induced by other product or algorithmic changes.

#### 3.2.7 Monitor outputs and outcomes

In practice, product teams will continually monitor the diversity of recommended items to detect operational failures (e.g., bugs or system failures), or (induced or exogenous) changes in the user distribution and user behavior. A survey which asks users or human raters to detect "productive" diversity may also be regularly employed to detect model drift and produce updated training data. Finally, determining whether the ultimate goal is being achieved, or is still worthy of being achieved, requires ongoing evaluation. This could involve survey methods to assess metrics such as affective polarization, and ethnographic research to understand what it means for users to be encouraged to be "more tolerant" in this way.

## 4. Value Measurement

In order for recommender systems to incorporate human values, there need to be methods for measuring how well a system is adhering to, promoting, or facilitating these values over time. This section describes how to go from a value to a metric, the issues involved in designing good metrics for values, and the data that is available on which to build such metrics.

Our description of recommender operation in section 3 didn't show what comes before and after recommendation proper. Figure 2 expands our view to include the process of selecting metrics, the interactions a user has, and the ultimate outcomes.

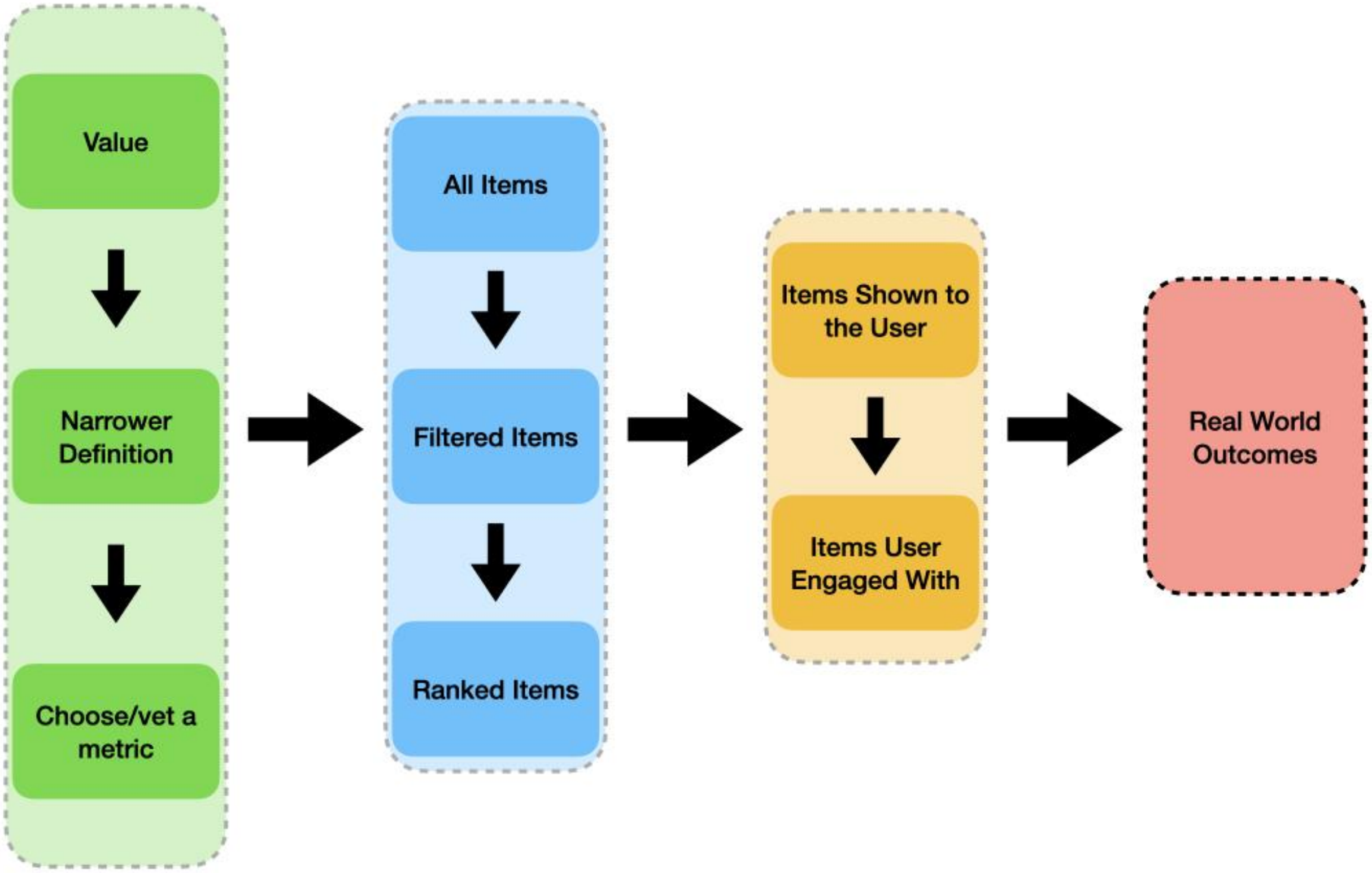

Figure 3: An illustration of the relationship between values, metrics, recommenders, items, and outcomes. The selection of a metric leads to a selection of items, some of which the user will engage with. Over time, the items a user engages with shape outcomes of interest.

## 4.1 From Value to Metric

Defining a metric is part of the process of specifying what, exactly, a particular human value means in the context of a real system. This process is called *operationalizing* a value. The people involved in defining metrics have considerable influence over the ultimate function of a recommender, which is why multi-stakeholder involvement in recommender metric selection may be important (Lee et al., 2019; Stray, 2020).

To illustrate the gap between a value and its measurement, consider the value of "safety," and in particular, protecting users from hate speech. A precise definition of hate speech is not only hard to articulate, but under constant debate and evolution, and every choice has some set of undesirable side effects (Díaz and Hecht-Felella, 2021; Gordon et al., 2021; Martinez and Paluck, 2020, 2020; Papcunová et al., 2021). Well-being is an even more complicated example. As discussed, well-being has many components (e.g., physical health, health of personal relationships, having purpose in life, etc.) and can be considered in the short term (e.g. entertaining content) or the longer-term (e.g. learning useful skills or fostering relationships). Making news recommendations that serve the public interest likewise requires defining concrete metrics that reflect some assessment of that interest (Fields et al., 2018; Vrijenhoek et al., 2020). At some point, a real recommender must commit to specific operationalizations of broad concepts, with the resulting tradeoffs between competing values and stakeholders.

## 4.2 Characteristics of a Good Value Measurement

A good measurement has a number of desirable properties, including validity and reliability (Mohajan, 2017), fairness, and legitimacy. Quinn et al. (2010) summarize the situation as:

> The evaluation of any measurement is generally based on its reliability (can it be repeated?) and validity (is it right?). Embedded within the complex notion of validity are interpretation (what does it mean?) and application (does it "work"?).
> (Quinn et al., 2010, p. 216)

Many social science theories involve quantities that are not directly observable and hence must be inferred, often with considerable uncertainty, which makes any measurement instrument implicitly a model of some purported underlying reality (Jackman, 2008). Construct validity is the convergence of a successful theoretical idea with a measurement that effectively captures it (Strauss and Smith, 2009). Jacobs and Wallach (2021) propose several other types of validity in the context of measurements within computational systems, including face validity (is the metric plausible?), content validity (does the metric capture all the relevant parts of the concept of interest?), convergent validity (does this metric agree with other accepted metrics?) and discriminant validity (does this capture something different than other metrics?). Reliability is typically evaluated in terms of agreement between multiple measurements (test-retest reliability), between different human judges (inter-rater reliability) and, for surveys, between different ways of asking a question (inter-item reliability) (Jackman, 2008).

It is rare that a particular measurement fully captures what we mean by some value in a particular context, meaning that most metrics are in fact proxies for what we actually care about. Even a good metric will change meaning when it is known to be used to make consequential decisions, e.g. student test scores must be interpreted differently when they are used to decide academic progression, because instructors will begin "teaching to the test." This effect is sometimes known as *Goodhart's law*, but there are a variety of different causal structures which can produce feedback processes that widen the distance between a metric and the underlying value (Manheim and Garrabrant, 2018; Mullainathan and Obermeyer, 2021). In the technical community this has been most discussed in the context of the general problem of algorithmic optimization and the difficulty of objective specification (D'Amour et al., 2020; Hadfield-Menell and Hadfield, 2019). Hence, as part of being precise about the definition of a human value, it is important to identify gaps in what is measured and monitor them over time. Human values — that is, what is considered important – also tend to change over time. Qualitative user research plays a critical ongoing role in designing and evaluating metrics.

Because modeling assumptions are required to connect a measurement to the underlying value it purports to reflect, measurement itself has fairness implications. The signals from measurements can vary across demographics, even after controlling for differences in user intent and task (Mehrotra et al., 2017). For example, if older users read more slowly than younger users, then a metric based on dwell time will be an over-optimistic measurement for older users regardless of their level of satisfaction. Thus, interpreting metrics at face value may systematically disadvantage and misrepresent certain demographics and user groups. Organizations deploying recommendation systems often rely on *internal auditing* methods as a way to measure how the overall system performs across different demographics and other user attributes (Feldman et al., 2015; Raji et al., 2020). External audits may also be required by regulation, as discussed below.

It is not enough to have an accurate and fair measurement of a human value. Stakeholders in the recommender system need to be able to accept the process and outcome of the measurement as legitimate. This means that two metrics that are operationally identical (e.g. they both correlate strongly with a desired outcome) may not be interchangeable. Transparency around how the measurement is carried out may help build trust in a metric. For example, some platforms periodically release public transparency reports which include various metrics (Singh and Doty, 2021b). Another way to increase legitimacy is to establish accountability regarding a measurement, e.g., through independent, external audits of the measurement (Suzor et al., 2018). More ambitiously, a metric could be created or chosen through a participatory process. For instance, the measurement could aggregate the opinions of a panel of users, as in the "digital juries" (Fan and Zhang, 2020) and "citizens assembly" concepts for making platform decisions (Ovadya, 2021). In one case, representatives of various stakeholders participated in the construction of a recommender that matched supermarket excess food donations to volunteer drivers and local food banks, using an elicitation process to define a ranking function (Lee et al., 2019).

### 4.3 Data Sources for Measuring Values

Our news recommender diversity example involved a complex, multi-step process for defining an algorithmic measure, ultimately drawing on human labeling to define the value of diversity. Broadly speaking, there are three main data sources for value measurements on a recommender platform.

The first source of data is the behavioral signals that users generate during normal usage, e.g. articles a user clicks on, songs a user plays, comments, emoji reactions, re-sharing of content, ads a user clicks on, purchases made, time spent on the platform and on specific items, and so on. These sorts of implicit signals of value are often called "engagement," but we note that some behavior signals are explicitly designed to give feedback to the algorithm, such as swipe left/right or thumbs up/down. Implicit and explicit behavioral feedback both provide distinct and useful information to a recommender system (Jawaheer et al., 2010).

The second source for signals of value is answers to survey questions that are posed to a fraction of the user base, typically a very small fraction. Surveys can ask very targeted questions, e.g., Facebook has asked users whether individual posts were "worth their time" (Facebook, 2019) while YouTube uses user satisfaction surveys that ask users to evaluate previous recommendations (Goodrow, 2021; Zhao et al., 2019). Survey results can be used to monitor real-world outcomes, evaluate A/B tests, and ultimately recommend items that a user is predicted to respond positively to when asked about their experience.

The third source for value measurements is human annotation. This data is often produced by paid raters, though human ratings also come from users flagging or reporting items. Vast amounts of human training data are routinely used to create models for identifying particular kinds of positive or negative content for the purpose of content moderation and ranking. Platforms that feature professionally-made content can also ask creators to provide metadata for a variety of ranking purposes. For instance, the Swedish public broadcaster asks editors to rate each story in terms of importance, public service value, and lifespan (Mccaffery, 2020). While survey data is limited by how often users can be asked to fill out a form, annotation data is costly to create and therefore available in limited quantities, and some types of labeling work may contribute to mental health issues for raters (Arsht and Etcovitch, 2018). Human annotation can also be noisy, inconsistent and biased depending on how rater pools are selected, and have limited, unbalanced coverage, while the meaning or usage of labels can change over time. Research in the area of human computation (Law and Ahn, 2011) attempts to address such issues (Faltings et al., 2014).

In contrast to surveys and annotation, the benefit of behavioral signals is that they are plentiful—many orders of magnitude more data is available. However, the behavior of users on the platform typically correlates with but does not perfectly capture any particular type of value, and is only a proxy for what different stakeholders actually care about. Moreover, behavioral signals have been shown to be sensitive to a variety of factors, such as the user and their demographics (Carterette et al., 2012; Hassan and White, 2013), the context in which the user is interacting with the system, and the recency of interactions. For example dwell time, a behavioral signal that has been used as a proxy for user satisfaction (Yi et al., 2014), can vary significantly depending on whether the item is the first one clicked in a list of results or not (Borisov et al., 2016). As discussed above, behavior does not represent underlying preferences for a variety of reasons including cognitive biases, information asymmetry, coercion, lowered expectations, and so on (Anderson, 2001; Bernheim, 2016; Camerer et al., 2004). Additionally, the collection of vast amounts of user data poses significant privacy concerns.

One important approach that tries to combine the benefits of behavioral and survey signals is to learn a model that predicts survey results given user behavior, that is, it tries to predict how a particular user would answer a survey question if shown a particular set of items. This method extrapolates limited survey data to all users, and is common in industry (Goodrow, 2021; Stray, 2020) but has not been much discussed. We view such prediction of survey results as an emerging method for aligning recommender systems with human values, though it must be understood that predictions are just proxies and must be continually monitored for divergence from ground-truth survey responses.

In principle, surveys can elicit complex judgements of value. In addition to the challenges of complicated, multi-component values (e.g., well-being) the context and wording of a survey can significantly affect the results (Stoutenborough et al., 2016; Weaver et al., 1997). Social desirability bias, the tendency of respondents to answer in ways that others would view favorably, adds a further complication (Furnham, 1986; Näher and Krumpal, 2012). Another issue arises when users from different demographic groups tend to answer questions differently (Eskildsen and Kristensen, 2011). Since online surveys often involve casual participants, “seriousness” checks and data denoising can be very important (Aust et al., 2013).

# 5. Design Approaches

We are now ready to discuss the major approaches that recommender system developers and researchers have used to promote the values discussed above, and in turn, to identify some of the challenges and open problems which require the development of new technologies. The structure of this section reflects the issues presented above. Section 5.1 discusses the core of the recommender system, namely the techniques used to select and order items shown to the user, and long-term user paths through these items. Section 5.2 discusses the affordances and controls made available to users and other stakeholders through UI and UX design. Finally, Section 5.3 considers fairness and multistakeholder perspectives, and describes techniques used by system designers to help optimize tradeoffs within real systems.

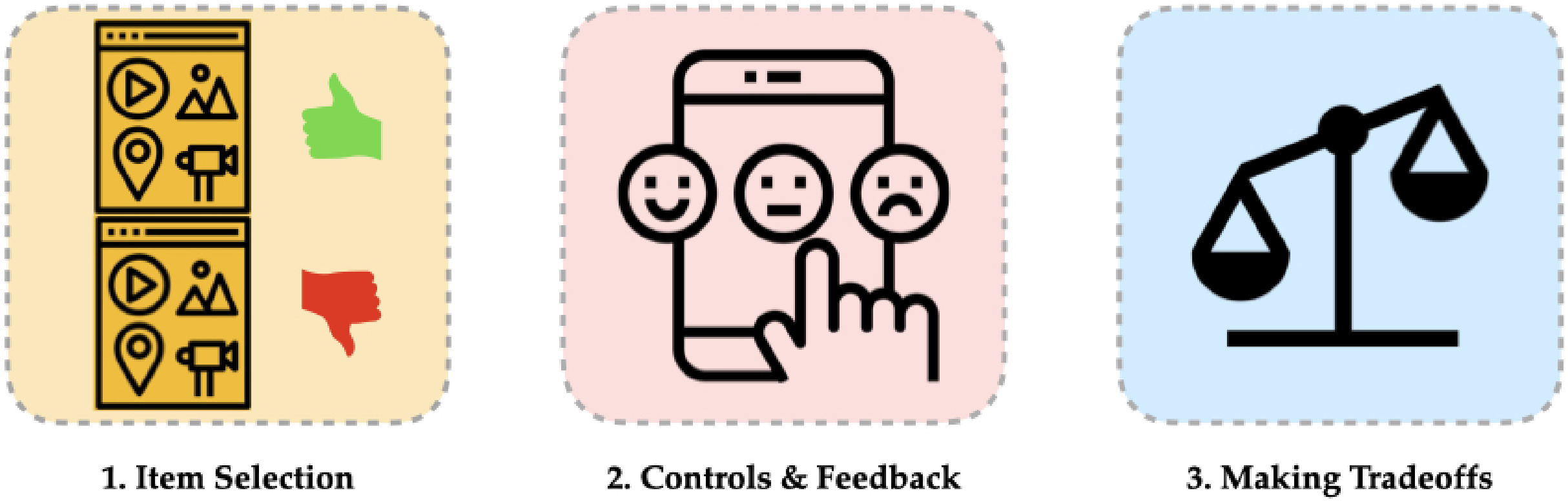

Figure 4: Section 5 discusses the design affordances that are typically used to modify recommendation systems to better align with values. We identify three primary options: 1) item selection, where designers intervene on the ranking of content to achieve a desired goal; 2) controls and feedback, which creates affordances for users to modify their experience on the platform; and, 3) by identifying and optimizes tradeoffs within the systems to better account for multi-stakeholder perspectives.

## 5.1 Item Selection

At the heart of any recommendation system are the models and algorithms used to generate one or more recommended items. Generally, the core models are trained to predict a user's behavioral response to a candidate recommendation (e.g., watch, like, reply, purchase) using properties of the user, the candidate items, and the context. Items are then scored by combining predicted outcomes in some way (Milli et al., 2020; Zhao et al., 2019). The nature of the properties used to make predictions, the predicted responses themselves, and the way these predictions are combined into a final score all play a role in the values that a recommender embodies.

### 5.1.1 Item ranking signals

The properties of an item being considered for recommendation—whether a social media post, news article or musical track—play a key role in determining whether it is of interest to the user, and therefore whether showing it to a user can serve one of their values.

The "topic" of an item is an important signal. A variety of techniques have been used for text analysis in recommender systems, including latent semantic indexing in Google News (Das et al., 2007), Latent Dirichlet allocation (Jelodar et al., 2019) and transformers. Image and video analysis are also used for topic assessment (Deldjoo et al., 2020), as is audio analysis (Yu et al., 2020). Modern recommender topic taxonomies can encompass tens of thousands, or even millions, of distinct categories and sub-categories. One of the challenges is that these classifiers are typically accurate for the popular elements of the taxonomy of topics, but often do not perform well for posts on less popular topics.

Topics do not directly correspond to specific user needs or to values. A post about "football" can be about organizing a football viewing party (possibly contributing to the value of connection) or just reporting the latest scores or player injuries (the value of knowledge). To support the value of empathy and care it would be useful to identify posts where the poster could benefit from support of their network (e.g., after losing a loved one, needing advice on a certain matter, or announcing an important event in their life). An open challenge is inferring the way a particular user will *relate* to an item, rather than analyzing properties of the item alone. There is nascent work on predicting the experience a user may have when viewing content, e.g., whether they are entertained, angry, or inspired (Chen et al., 2014; Ignat et al., 2021). These sorts of inferences could also enable better targeted or more persuasive advertising, which raises policy concerns (Singh, 2020b).

Content analysis is also heavily used by social media platforms in order to determine whether a post violates community policies or is of low quality in some way, and should therefore be removed or demoted. The predominant method for determining such violations uses ML models trained on content labels provided by paid raters. These models increasingly use multimodal techniques, simultaneously considering text, speech/audio, images and video. This is often crucial to understanding the intent of a post, e.g., the caption "love the way you smell today" means something different when superimposed over the image of a rose vs. a skunk (Kiela et al., 2021).

Increasingly, items are evaluated not just in terms of their content but their context, including properties of the poster and audience, previous user behavior, reactions from other users, and so on (Halevy et al., 2020). Incorporating embeddings of sequences of user interactions has been shown to increase accuracy in misinformation and hate speech classification on Facebook (Noorshams et al., 2020) while "toxic" conversations on Twitter are better predicted by including network structure features (Saveski et al., 2021).

These models are generally trained on labels or annotations generated by human raters. This process itself is potentially subject to error and particular forms of bias, both in the instructions given to evaluators and their assessment biases (Sap et al., 2021). Also, many of these categories are both complex and politically contested (Bozarth et al., 2020; Martinez and Paluck, 2020) and raters often disagree, which complicates the evaluation of model accuracy (Gordon et al., 2021). The promotion or demotion of particular topics or types of language, and the errors and biases in this process, have major implications for freedom of expression, which implies a deep connection between these technical methods and broader policy considerations (Douek, 2021; Keller, 2018; York and Zuckerman, 2019).

#### 5.1.2 User Trajectories

We use "trajectory" to refer both to the sequence of items a user saw over time and the reactions those items evoked. A user's past trajectory often provides information that can be used to make better predictions about their future preferences and behaviors. At the same time, we want to ensure that the user's future trajectory supports the values they care about. Moreover, trajectories are the basic unit for studying some problems users experience on platforms, as they are central to discussions about potential long term effects on, say, well-being and polarization. For example, the "filter bubble" critique is essentially a statement about the typical course of user trajectories.

The majority of recommenders deployed in practice are "myopic" in the sense that they make predictions of a user's immediate response to the next slate of items presented, and rank items based on these predictions. However, many recommenders use past sequences of user and system behavior in sophisticated ways.

Advances in deep learning have made possible the practical deployment of *sequence models*, such as recurrent neural networks or transformers. For example, Beutel et al. (2018) describe a recurrent (specifically, an LSTM) model deployed at YouTube. Recurrent approaches explicitly model a user's "latent" or "hidden" state, that is, they include variables which represent aspects of the user's situation or psychological state which are unobserved but have effects on what the user wants to see next. This state might encode aspects such as user satisfaction, frustration, or current topic focus, but interpretation of the hidden state embodied in such models is challenging and is tightly coupled to the engagement metrics being predicted and optimized. Inferring user state from behavior is an important challenge and a key step toward better supporting many values. For example, a recommender could detect a user's dissatisfaction with a certain type of content, or with too little topic diversity in the recommendation stream, or even that someone was developing an eating disorder (Yan et al., 2019).

Current ranking techniques do not offer the means to (directly) optimize a user's *future* trajectory. A promising direction is the use of *reinforcement learning (RL)* for optimizing such futures non-myopically (Afsar et al., 2021; Shani et al., 2005; Taghipour et al., 2007). In particular, the use of RL allows the system to consider the impact on the user of not just immediate recommendations, but of the entire sequence of recommendations it might make over an extended period, and plan that sequence accordingly.

In our news diversity example, the redesigned recommender considered diversity within each slate of recommended items independently. An RL-based recommender would be able to consider the diversity of items over days or weeks as well. In an educational setting—where a user's understanding of a topic is best served by a graduated but individualized and adaptive sequence of content—RL can be used to plan and adapt the appropriate recommendation sequence. As such, RL offers considerable promise as a technology for individual or user-level value alignment. To date it has been used largely to optimize user engagement over the long term, but with suitable metrics and objective criteria, it could play a vital role in better aligning recommendation trajectories with user well-being by adaptively planning recommendation sequences.

As with sequence models, advances in deep RL methods have made it more practical to deploy RL in practical recommender systems (Chen et al., 2019; Gauci et al., 2019; Ie et al., 2019). That said, a number of challenges remain. First, since RL relies on sequential interaction data with real users, such models are often trained offline using data generated by past recommendations, which may induce bias because the user was interacting with a different model (Chen et al., 2019). The second challenge involves choosing from a large action space ranging from hundreds to millions depending on the recommendation task (Dulac-Arnold et al., 2015; Ie et al., 2019). A related problem is that item recommendations are often made jointly in slates or scrolling feeds, where the interactions or interference among the visible items makes interpreting and optimizing for user responses challenging (Ie et al., 2019). Finally, adopting RL for true value-alignment requires sophisticated models of various aspects of user latent state (e.g., preferences, satisfaction, awareness/knowledge levels, fatigue/boredom, etc.) and their dynamics. These psychological and situational states are challenging to uncover from observable user-response behavior, and may require planning over extremely long "event horizons" as user adaptation to recommender changes may take 3-6 months to fully materialize (Carroll et al., 2021; Mladenov et al., 2019).

Some work has considered trajectories that might be unhealthy or harmful. One problem concerns minimizing the likelihood that trajectories will end up with users viewing large amounts of questionable content, such as videos that might promote unhealthy behavior, false statements about COVID vaccines (Chan et al., 2021), or other content that may not explicitly violate the platform's policies. This might occur not because users explicitly ask for such material, but because they end up in places (e.g., groups) where this material is common. In addition to downranking such content, another solution is to avoid recommending low quality groups or content sources to users.

Another strand of work concerns items that are acceptable or appropriate when considered in isolation, but could be harmful if consumed too much or by certain vulnerable people. For example, there may be nothing wrong with a diet video but perhaps someone with an eating disorder should not be presented with an endless stream of such videos; or it may be unhealthy to watch exclusively violent movies. Singh et al. (2021) address this possibility by ranking user trajectories by proportion of "unhealthy" content, then using the mean proportion of this content in the top $\alpha$th percentile of user trajectories as a regularization metric. They demonstrate a safe RL approach that improves both worst-case and average-case outcomes, in terms of the fraction of "unhealthy" items recommended to any one user.

The most complex concern about trajectories is the possibility that recommender systems might change user preferences in a manner that increases engagement but harms some other value. This is commonly associated with the idea of "manipulation," meaning unwanted changes in attitude or behavior (even if unintentional).

This sort of optimization-driven shift has been frequently suggested as a mechanism driving filter bubble, echo chamber, or polarization effects, though the empirical evidence is mixed (Bruns, 2019; Guess et al., 2018; Stray, 2021a; Zuiderveen Borgesius et al., 2016). Such models posit a feedback loop where users choose particular items (as in selective exposure effects (Prior, 2013)) and the recommender responds to that engagement by narrowing its output to those topics, which in turn shifts user preferences further in that direction. This effect may be a particular concern for RL systems, which may learn how to make their users more predictable so as to maximize engagement (Russell, 2019). A polarizing preference shift effect has been demonstrated in multiple simulations with different specifications (Carroll et al., 2021; Evans and Kasirzadeh, 2021; Jiang et al., 2019; Kalimeris et al., 2021; Krueger et al., 2020; Rychwalska and Roszczyńska-Kurasińska, 2018) which suggests that it could be a robust effect.

A range of work has attempted to determine the causal effect of recommender systems on content consumption trajectories. This is particularly important if those trajectories are correlated with offline outcomes such as violence (Global Internet Forum to Counter Terrorism, 2021). There is replicated evidence that strongly moralizing expression spreads faster than other content on social networks (Brady and Van Bavel, 2021) and such moralizing seems to precede offline violence (Mooijman et al., 2018). A number of researchers interested in categories such as "far right," "conspiracy," or "radical" have studied the network structure of recommendations between YouTube channels (Faddoul et al., 2020; Ledwich and Zaitsev, 2019; Ribeiro et al., 2020a). While this showed that more extreme channels often link to each other, these studies do not analyze user trajectories because they were conducted without any personalization. A different approach is to program bots to selectively engage with certain topics. This has generally shown that engaging with some type of content increases its frequency (Wall Street Journal, 2021; Whittaker et al., 2021) but this design models users as unchanging, so it does not provide evidence on persuasive effects. A user-tracking study of "far right" content on YouTube from 2016 to 2019 found that consumption there matched broader patterns across the web, including consumption from sources without recommender systems (Hosseinmardi et al., 2020). Separating consumption due to recommenders and consumption due to users' seeking behavior is difficult without on-platform experiments. Using a set of Twitter users randomly assigned to receive a baseline chronological feed, Huszár et al. (2021) found that Twitter's home timeline recommender reduced the consumption of more politically extreme material.

In general, inference of the causal effects of recommenders on user preferences and outcome is a major open problem. The feedback effects between algorithms and societies are at the cutting edge of social science research, in need of both interdisciplinary and cross-sector collaboration.

### 5.2 Controls and Feedback

Meaningfully supporting human values requires effective communication between the user and the system, beyond the standard implicit signals (clicks, shares, dwell time, etc.). We place communication affordances on a continuous spectrum between explicit and implicit. We use controls to refer to features where the user can explicitly change certain settings, and feedback to refer to situations where the user is more passive and their preferences are elicited in some way, such as by answering a survey question, providing a like/dislike, etc.

There are a variety of documented benefits to providing users with more control over their experience. Users have reported being more satisfied with their recommendations when given greater control (Jin et al., 2017) and better controls have led to increased engagement (Jannach et al., 2017). Interestingly, even the appearance of controls can achieve some of these effects. In (Vaccaro et al., 2018), users reported higher satisfaction with the system even though the controls had random effects. Conversely, Loecherbach (2021) gave users control over the diversity of items in their feed but there was no correlation between actual and perceived diversity. In general, greater recommender control usage does not always translate into better control (Waern, 2004).

Existing systems offer a variety of ways for a user to customize their experience, and many more control and feedback schemes have been proposed by researchers (He et al., 2016). Some involve simple direct feedback on individual items, e.g., thumbs up / thumbs down, or item-level ratings (Boutilier et al., 2003; Zhao et al., 2013). Others involve evaluating the relevant importance of pairs of items (Boutilier, 2013; Chen and Pu, 2004). Even when controls are provided, many users do not know that they exist or what they do (Hsu et al., 2020; Singh, 2021), find them challenging to use, or simply don't see the value in engaging with them. As a result, most users do not use recommender controls and a "passive" user experience remains the default (Jin et al., 2017). Aside from better educating users on controls, a promising approach is to design feedback mechanisms that serve multiple purposes: such as sending a social signal (which is why the user would use it) and simultaneously providing direct feedback that pertains to some value. Some examples of this exist today, such as the "Insightful" emoji on LinkedIn, the "Care" emoji on Facebook, and the proposed Respect button (Stroud et al., 2017).

One of the issues that limits the use of controls is the difficulty in creating a direct, understandable link between the input a user provides and the resulting change in the system's behavior. In some cases this link is quite direct (e.g., "don't show me any sports-related content"), whereas in others the outcome of a control/feedback action is less obvious (e.g., "show me more videos like this video I just watched"). There are many points along this spectrum. For example, instructing the system to "show less about football" will be largely predictable in terms of the content it affects (though what about political statements/protests by athletes?) but users may still not understand the expected magnitude of change. To complicate things further, there is often some ambiguity about whether the change the control offers is transient (e.g., for the current session) or for the longer term. One proposed design is to highlight items that will be added and removed from the user's feed when a control is changed (Schnabel et al., 2020).

Organizing content into "channels" can help users to better customize their experience. Some systems include implicit ways of specifying what sort of items are to be included, such as music recommenders which can continue a human-generated playlist (Zamani et al., 2019). Algorithmic channels could also be designed around particular goals (e.g., learning to play guitar) or needs (e.g., getting support from friends, participating in lively conversations). The primary technical challenge here is building relevance measures that capture the dimensions that users care about.

In the future, communication between the user and the system will take different forms. Conversational recommender systems are an emerging area that offer a potentially more natural, engaging, and usable interface for people to express their preferences, particularly in light of recent advances in speech, natural language and dialog technologies (Adiwardana et al., 2020). Applications range from integrations with the approaches listed above to entire reimaginings of the control process (Christakopoulou et al., 2016; Jannach et al., 2021). A related idea is recommender personae, where the recommender is associated with a particular personality (e.g., explorer, diplomat, expert) to set a particular context for the conversation (Harambam et al., 2018). Conversational recommenders face many of the same challenges as other control systems, but moreso: they must interpret the metaphorical, imprecise, or subjective language a user may use to convey their needs or topic preferences.

#### 5.2.1 Meaningful explainability

The field of explainable recommender systems has grown into a vibrant research area (Zhang and Chen, 2020) in part because users respond positively to having good justifications for why certain items are suggested. Tintarev and Masthoff (2015) argue that a good explanation can increase transparency, scrutability, trust, effectiveness, efficiency, persuasiveness and satisfaction in the system. Each of these terms expresses somewhat different goals: transparency shows how the system works and can be instrumental in accountability; scrutability allows the users to tell the system it is wrong; trust increases confidence in the system; and satisfaction increases users' sense of derived benefit; effectiveness helps the user employ the system toward their own aims; persuasiveness convinces the user to change their beliefs or behavior; efficiency can increase the value of the system. Some of these values can be contradictory and may not be achieved at the same time. For example, an explanation that increases transparency of the system does not necessarily increase trust if the explanation is not understandable or reveals undesirable behavior. Therefore, what constitutes a "good" explanation very much depends on the goal, and the field is rapidly developing (Hase and Bansal, 2020; Narayanan et al., 2018; Nunes and Jannach, 2017).

With the rise of increasingly complex machine learning models, it has also become increasingly difficult to give intuitive and understandable explanations of why a user received a specific recommendation. Explanations may be shown to the user in different forms (e.g., text, visuals, highlighting relevant features) and may either attempt to explain the workings of the recommendation model itself, or may be the result of training a separate model that generates post-hoc explanations from model inputs and outputs (Zhang and Chen, 2020). There are also recommender algorithms specifically designed to be explainable (Xian et al., 2021). The system of Balog et al. (2019), for instance, operates based on set intersections, e.g., "recommended for you because you don't like science fiction movies except those about space exploration." This explainable-by-design approach avoids the challenges of interpreting learned models (Rudin, 2019), but generating understandable explanations from deep learning models is an active area of research that may yet prove fruitful (Kim et al., 2018; Kokhlikyan et al., 2020; Sundararajan et al., 2017).

### 5.3 Making Tradeoffs

Our news diversity example involved a tradeoff between a diversity metric and a short-term engagement metric. In general, there are numerous tradeoffs involved when incorporating human values into recommender systems, and a variety of techniques to evaluate these tradeoffs and make choices. There are at least three categories of tradeoffs: 1) tradeoffs between the benefits and potential harms to different people, 2) tradeoffs between different types of stakeholders, such as content creators versus content consumers, and 3) tradeoffs between values, resulting in different (but not obviously worse or better) measurable outcomes induced by various recommendation strategies.

Ultimately there must be some notion of a fair or optimal tradeoff, and techniques for making "good" tradeoffs in this sense. We hope that these tradeoffs are informed by the expressed opinions of users and other stakeholders, so we first discuss the theory of social choice, which studies how to combine the preferences of many people. We then discuss techniques for achieving various notions of fairness, such as between different types of users or between different stakeholder categories. Finally, we discuss tools designed to optimize tradeoffs when faced with the practical necessity of tuning a recommender's ranking function.

### 5.3.1 Tradeoffs in theory: Social Choice

People express their values in their everyday use of recommender systems. Indeed, many recommender controls are designed specifically for this purpose, everything from upvoting and swiping left/right to reporting violating content. This creates the problem of aggregating preferences, and negotiating between competing desires of individuals, groups, and stakeholders. The framework of *social choice,* originally developed in economics (Gaertner, 2009; Sen, 2017), provides a foundational tool for addressing tradeoffs at all of these levels.

Abstractly, this framework assumes that each stakeholder has a *utility function* over a set of possible outcomes for them. This is motivated by the idea that someone could say which of several situations they'd prefer, that is, that each person has *preferences*. Note that this utility function can be purely "local" (a user may care only about whether they get good recommendations) or it can involve societal values (a user may care that other users also like what they like, or that vendors are treated fairly). A *social welfare function* is a voting process or a way of "adding up" or combining stakeholder preferences to produce a single number, a "societal utility" or social welfare. The aim is then to adopt a recommender policy which maximizes the expected social welfare.

This formulation can encompass virtually any criteria that express preferences over short or long-term, individual or group outcomes. There are relatively direct mathematical expressions for penalizing addictive behaviors, group-level diversity of consumed content, fairness across individuals, and so on. Conversely, any collaborative recommender system aggregates the explicit or implicit preferences of users in some mechanical way, as when signals like upvoting and watch time are combined across many users to decide how items are ranked for a different user. Social choice theory is a key bridge between the normative and the algorithmic, useful for analysis and design.

Although social choice theory is foundational, there are two major difficulties to applying it in practice. First, determining what an individual's "preferences" actually are is quite challenging both in theory and in practice (Thorburn et al., 2022a). There is a long history of formal elicitation methods such as asking users to repeatedly say which of two options they prefer, or asking them to play various kinds of economic games (Camerer and Fehr, 2004; Chen and Pu, 2004), but formal preference elicitation can be involved and is quite demanding of the participants. Further, someone might be uninformed, coerced, addicted, have lowered expectations (Anderson, 2001) or want something that isn't available (Freedman et al., 2020; Robertson and Salehi, 2020). In addition, attempts to elicit preferences can lead to strategic behavior where people misrepresent their beliefs to try to induce more favorable recommender outcomes, e.g. a content creator usually has an incentive to argue that their content is relevant to as many users as possible (Ben-Porat et al., 2019; Ben-Porat and Tennenholtz, 2018). While many different kinds of feedback can provide crucial signals of what people value, behavior cannot be naively interpreted as true preferences.

Second, there is no entirely bottom-up or value-neutral method of ethics. Simply specifying the outcomes over which stakeholder preferences are defined is inextricably tied to the values being considered (Bernheim, 2016). For example, there is the choice of what will be voted upon. There is also the question of whose preferences count in what situations, e.g. community administrators may have special voting privileges, and it may be important to encode "rights" that cannot be infringed by the preferences of others. Hence, social choice approaches cannot excuse system designers from making consequential normative choices (Baum, 2020). There is more to democracy than voting systems.

### 5.3.2 Fairness and Tradeoffs among Stakeholders

Many of the values in our list involve making trade offs between different stakeholders (such as users vs. content providers) or among members of a stakeholder group (such as different subgroups of content providers which compete for attention). Correspondingly, there are a wide variety of notions of "fairness," which often (but not always) are framed as some sort of trade off. The extent to which these tradeoffs are inherent is an open question in the research literature, because there are cases where it is possible to improve performance for one user subgroup without decreasing performance for other users. There are multiple challenges in this area, including defining fairness, measuring it in practice, and designing algorithms for efficient recommendation. See Ekstrand et al. (2021) for an overview.

Several major categories of fairness have been proposed in the context of recommender systems, roughly corresponding to the interests of different stakeholders. "C fairness" considers how well individual information consumers are served, "P fairness" is concerned with the distribution of attention between items or providers of content, while "CP fairness" considers both simultaneously, as in a rental property recommender designed to protect the rights of both minority renters and minority landlords (Burke, 2017; Wang and Joachims, 2021).

Recommender systems are mostly evaluated based on average performance across all users, but different user subgroups, such as age or gender groups, might be served with differing performance or error rates. Subgroup performance disparities can happen for a variety of reasons, including differences in group size or activity that affect the amount of training data available (Ekstrand et al., 2018; Li et al., 2021). There is a large body of work on mitigating group-level unfairness in classifier models, some of which can be adapted to recommender systems. For example, (Beutel et al., 2019a, 2019b) use pairwise comparisons of the ranking of different items to generalize the well-known "equality of opportunity" and "equality of odds" measures, showing that it is possible to equalize prediction error rates between user groups on a large commercial platform. However, algorithmic approaches that aim to equalize effectiveness disparities between user groups may make inappropriate tradeoffs: increasing recommendation utility for one user does not necessarily require decreasing it for other users, so it is not clear that allowing a solution that may decrease utility for well-served users is appropriate, as opposed to other approaches such as feature prioritization in the engineering process (Ekstrand et al., 2021 §5.4).

Items and their producers, on the other hand, necessarily compete for user attention. This leads to the concept of "exposure fairness" which may be formulated in a variety of ways (Singh and Joachims, 2018). Even if predicted "relevance" scores correctly measure the value of an item to an individual user, slightly less relevant items may get disproportionately less attention simply because they appear farther down, an effect known as "position bias." Several algorithms have been proposed to ensure item attention is proportional to item value on average, either across a group of items or across multiple slates of recommendations (Biega et al., 2018; Diaz et al., 2020; Wang and Joachims, 2021). Such algorithms can be considered a correction to a type of error or "technical bias" in the fair ranking typology of Zehlike *et al.* (2021). More normative definitions of item fairness are often desirable. For example, Spotify strives to give exposure to less popular artists to counteract the "superstar economics" of cultural production (Mehrotra et al., 2018), while a "demographic parity" conception of fairness may be appropriate when qualified candidates from different groups (say, men and women) should be shown to prospective employers at the same rate (Geyik et al., 2019). A wide variety of fairness metrics concerning the exposure of items, groups of items, or producers have been proposed, though many of these are closely related (Kuhlman et al., 2021; Raj and Ekstrand, 2022; Sapiezynski et al., 2019; Zehlike et al., 2022).

Where it is possible to produce reasonable quantitative estimates of utility to different stakeholders, *multi-objective optimization (MOO)* can be used to balance multiple conflicting stakeholder utility and fairness objectives. One approach is to ensure that the recommender is Pareto efficient, meaning that there should be no way to modify a slate of recommendations to make it more fair without reducing utility, or to increase utility without reducing fairness (H. Wu et al., 2021). Mladenov et al. (2020) go beyond this by proposing a recommendation method that maximizes user social welfare (total user utility) by allowing small sacrifices in utility from well-served users to drive large gains for less well-served users. Reinforcement learning has also been applied to multi-sided fairness, through contextual multi-armed bandits which simultaneously optimize stakeholder utility and fairness objectives (Mehrotra et al., 2020, 2018; Wu et al., 2022). Recently, several researchers have taken a game-theoretic approach to the study of recommender systems. Ben-Porat and Tennenholtz (2018) and Ben-Porat et al. (2019) develop approaches that account for the strategic behavior of content providers while aiming to maximize user engagement. While all these methods hold promise for value alignment in complex ecosystems, they have not seen practical deployment to date.

### 5.3.3 Optimizing Tradeoffs

Because there is no purely "bottom up" way of making tradeoffs, system designers must ultimately choose some set of overall objectives or "ground truth" signals to serve as overall measures of value. Increasingly, AI tools are used to help make tradeoffs over the complex design space of recommender parameters, particularly the relative "weighting" of the signals that feed into item ranking functions.

Milli et al. (2020) determined the relative value of different user actions including viewing a Tweet, sharing it, liking it, etc. by connecting these interactions to the sparse use of the "see less often" control in a causal Bayesian network. This network represents dependencies such as the fact that a user has to view a Tweet before they can share it. By taking "see less often" as a ground truth signal of negative value it was possible to infer the value of all of the other, more common interactions. This idea generalizes to more complex methods for finding multiple weights that optimize multiple metrics.

Bayesian optimization (Golovin et al., 2017; Rasmussen and Williams, 2006) can be used to find the weights of a ranking function that maximize relevant (perhaps long-term) metrics, and automatically run data-gathering experiments to improve those predictions (Gupta and Ouyang, 2020). This approach requires that the designer be able to assess the overall "utility" of any vector of performance metrics, which itself induces various tradeoffs. For example, is a product that has a greater number of items viewed but less total time spent and lower reported satisfaction better than the opposite? The tools of interactive optimization and utility elicitation (Boutilier et al., 2003; Zhao et al., 2013) could play an important role here, though these approaches have not yet found widespread use in practical recommender design.

## 6. Policy Approaches

In the previous section we have discussed design approaches to recommender systems and human values, that is, potential product changes. This section considers policy-making, which can be a powerful lever for change. Policy-making is informed by all of the perspectives articulated above, including ethical, procedural, measurement, and technical issues. Chosen policies impose constraints on how to build a recommender system and introduce additional technical challenges.

By policy-making we mean external governance, from government, regulators, or external bodies with appropriate authority. All large platforms have internal policies as well, particularly around which types of items are eligible for recommendation, but we focus on external governance as an important interface between recommender system operators and the rest of society. Because recommender systems are used in so many different types of products, we do not offer specific policy recommendations. Instead, our goal is to discuss the major categories of policies that have been proposed, and especially to understand how these policies could be translated into the terms of metrics and algorithms. At the current time there is a large gap in terminology and understanding between the recommender technology and policy communities, which we seek to highlight and begin to address. There are also policy-relevant technology gaps: the capability to do what policy-makers ask may not yet exist, as in the "right to explanation" provisions of the GDPR (Wachter et al., 2017).

We consider policy approaches that are relevant to recommender systems specifically, as opposed to social media or online platforms generally (neither of which necessarily involve recommender systems). We do not discuss content moderation policy approaches here, but direct readers to reviews such as (Douek, 2021; Keller, 2018; York and Zuckerman, 2019).

### 6.1 Risk Management vs. Value Sensitive Design

One proposed policy approach would require recommender system operators to evaluate the potential risk of harm from operating their systems. This is the approach taken by the proposed EU Digital Services Act, which requires "very large online platforms" (currently defined as those with more than 45 million users in the EU) to perform yearly assessments for three kinds of risks: the dissemination of illegal content, negative effects on fundamental rights, and "manipulation" with effects on "public health, minors, civic discourse, or actual or foreseeable effects related to electoral processes and public security." (European Commission, 2020) Any harms found must be mitigated through various means including

“adapting content moderation or recommender systems.” The proposed Digital Services Oversight and Safety Act of 2022 (DSOSA) in the U.S. takes a similar approach, requiring platforms of certain sizes and scope to conduct assessments and mitigate any risks, including by “adapting the content moderation or recommender systems (including policies and enforcement) of the provider.

This touches on several of the values in our table, but is relatively narrow in two ways. First, a risk-based framework is concerned only with potential harms. Second, the type of risk mitigation envisioned by the DSA and the DSOSA generally happens after a system is already built. Another approach is to require consideration of important values during the design phase, as with “privacy by design” provisions (Drev and Delak, 2021). Extending this to more general values, one German law requires platforms to meet certain content diversity obligations (Helberger et al., 2019).

The challenge for policy-makers or regulators is to be both precise and general about how harms are to be assessed and values are to be enacted in recommender systems. In principle this could involve monitoring certain metrics, as is already done in environmental regulation and media monopoly policy. Such regulation would face all of the challenges of choosing metrics discussed above, and certainly no one set of metrics will be appropriate for all recommender systems. Even if useful metrics for harm or good can be found, there is the difficult question of what constitutes an “acceptable” value (Nechushtai and Lewis, 2019).

## 6.2 Accountability

Policy approaches to the issue of *accountability* include provisions around transparency and evaluation mechanisms such as audits. Both are instrumental in supporting other values such as agency, control, and accountability.

Transparency has been a major focal point of legislative efforts in the United States and the European Union. Over the past two years, lawmakers in the United States have introduced numerous bills that seek to compel internet platforms to provide more transparency around how they develop and deploy algorithmic systems for content curation purposes (Mozilla, 2021; Singh, 2020a). The key challenge from a policy perspective is in defining the goal of transparency efforts, who they are meant to serve, and what, exactly, should be disclosed. Transparency may be intended to serve users, lawmakers, researchers, journalists, etc. (van Drunen et al., 2019). There are also privacy, security, and intellectual property issues that complicate disclosure (OECD, 2019, chap. 4; Shapiro et al., 2021; Singh and Doty, 2021a). Suggestions for what to publish have included the recommender source code, data on users (e.g. demographic data and their interactions with the system), options for users to modify recommendation “parameters”, data used for training models, key metrics, and the rationale behind product features and ranking changes (Stray, 2021b; van Drunen et al., 2019; Vorm and Miller, 2018). Each of these has limitations.

Production recommender code is extremely large, difficult for outsiders and non-technical individuals to understand, and may not be particularly revealing without reference to the content and user data it operates on. For example, user interactions over time can be used to understand people’s trajectories through the system, as discussed above. However, user data is difficult to share because of privacy concerns. This leads to the idea of sharing aggregated data, and many platforms already do so in the context of content moderation and targeted advertising (Singh, 2020b; Singh and Doty, 2021b). Then the key policy question becomes which metrics should be disclosed and how they are defined. While recommender operators will have key insights into what is relevant to measure, relying on them to select what is shared may pose a conflict of interest. Note that even aggregated user data can be used to re-identify individuals so it may need to be protected with techniques such as differential privacy (Dwork and Roth, 2014; Narayanan et al., 2016).

A related suggestion is that recommender operators should share the policies used to guide ranking and recommendation efforts, including what types of low quality or harmful content a platform downranks. The policy community has pushed recommender operators to disclose the rationale behind ongoing product changes more generally, including changes to recommender algorithms and parameters (Singh, 2020c; Stray, 2021b). This leads to the idea of a "change log" or "proceedings" that details what the operators were trying to do with each change (e.g., increase news quality), what data they had in front of them (e.g., fraction of items from each news source rated false by fact checkers), and what change they made (e.g., downrank certain sources by a certain amount). This is especially important as metrics or algorithms alone will not tell the full story: the motivation and context of a change are relevant to values. Thus far, a handful of civil society organizations have pushed platforms to adopt change logs for recommender system-related policies (Singh, 2020c).

Algorithmic auditing of AI/ML based systems has recently gained increased attention, and has mostly focussed on fairness and discrimination concerns in decision-making systems or predictive models (Vecchione et al., 2021). Although fairness remains a concern in recommender systems, in principle recommenders might be audited for any of the values discussed in this paper. A recent review of 62 academic algorithmic audits identified eight audits of recommender systems (Bandy, 2021), of which seven were concerned with "distortion" effects such as echo chambers or lack of source diversity. Several of these audits looked for, but did not find, echo chamber or filter bubble effects on Google News, the Facebook News Feed, and Apple News (Bandy, 2021, p. 16). They did find that a small number of news sources make up a large percentage of the results in Google News and Apple News.

While companies can and do perform internal algorithm audits of various types (Chowdhury, 2021; Raji et al., 2020) regulation could require external audits of recommender systems to mitigate these concerns (Singh and Doty, 2021a). Regulation could also direct audits to evaluate specific biases or harms to users or consumers (Brundage et al., 2020). The draft Digital Services Act requires certain platforms to undertake yearly audits to ensure they are meeting their risk assessment and mitigation obligations (European Commission, 2020). Similarly the Algorithmic Fairness Act in the United States would require covered entities to conduct and retain a five-year audit trail which includes information on how an algorithmic system was developed, trained, and tested (Singh, 2020a).

The personalized nature of recommender systems complicates auditing. Consider the problem of auditing user trajectories through recommender systems. Ribeiro et al. (2020b) collected data on non-personalized recommendations across YouTube channels and then evaluated trajectories using random walks, while the Wall Street Journal used 100 TikTok bots pre-configured to watch only videos on particular topics (Wall Street Journal, 2021). Neither of these methods model real user behavior. Conversely, Hosseinmardi et al.

(2020) used panel web browsing data collected by Nielsen to evaluate consumption of YouTube videos across the political spectrum, and The Markup's Citizen Browser project asks users to install a browser extension that reports what they see on Facebook and YouTube (The Markup, 2020). These observational studies have the advantage of involving real users, but the methods used so far cannot produce robust causal inferences about recommender effects. For example, it is currently not clear if social media contributes to depression or if depressed people spend more time on social media, or both. If these methodological issues cannot be solved, even extensive platform data sharing may not be sufficient to answer questions of interest. In that case, on-platform experiments may be the only reliable approach to auditing the effects of recommender systems, which would require extensive collaboration between recommender system operators and external researchers.

The nature of platform access remains to be defined. Industry players assert they face constraints when participating in third party audits, including concerns around privacy, security, intellectual property, competitiveness, and cost (OECD, 2019, chap. 4; Shapiro et al., 2021; Singh and Doty, 2021a). Many of these concerns are also risks to a third-party auditor, who must protect shared personal data yet typically do not possess the security resources of a platform. Additionally, many third-party auditing entities do not have the necessary technical skill and resources to audit recommender systems at scale (Singh and Doty, 2021a).

## 6.3 Translating Between Policy and Technology

There has long been miscommunication between the builders and regulators of technology. At the present moment, many countries are drafting or passing laws that regulate recommender systems of various kinds, especially social media, news recommenders, or targeted advertising. Unfortunately, much of the policy discussion taking place uses terms that do not map easily to recommender technical affordances (Singh, 2020a).

For example, Article 29 of the proposed Digital Services Act requires recommender-based platforms to disclose "the main parameters used in their recommender systems" as well as any available controls to "modify or influence those main parameters." (European Commission, 2020) Unfortunately, it's not clear what a "main parameter" is (Helberger et al., 2021). While this probably doesn't refer to the millions or billions of learned neural network weights, real recommender systems involve hundreds of major interacting components which are continually configured and tuned in complex ways. The vast majority of such internally configurable settings will not be understandable or useful to users, auditors or regulators. A better formulation might arise from considering the design of recommender controls, as discussed above.

The same provision also stipulates that recommenders offer "at least one option which is not based on profiling." Profiling in this context is defined in the GDPR and includes "personal preferences" and "interests" (European Commission, 2016). Thus, exclusion of profiling in this sense may exclude even Twitter's classic reverse chronological timeline, which cannot operate without the user indicating who to follow. An alternative approach would be to require an option that personalizes based on explicit user controls only, as opposed to implicit signals such as clicks or watch time.

"Amplification" is another word which appears in several proposed laws yet is difficult to translate into operational terms. Modern platforms can provide huge distribution to an item in a short period of time via information cascades, which result from a combination of user sharing and algorithmic recommendation (Bartal et al., 2020). While "amplification" is a reasonable name for this phenomenon, many definitions of "amplified" collapse to "shown" when parsed carefully, e.g. "the promotion of certain types of extreme content at the expense of more moderate viewpoints" (Whittaker et al., 2021). The selection of any type of content for display reduces the promotion of all other types because of the zero-sum nature of selection, so amplification would mean any display of "extreme" content when other content was available.

Other definitions are more specific, and compare the distribution of an item to some (often implicit) baseline. One approach is to define amplification as the prevalence of some type of content in user feeds relative to the prevalence of that type among all available content. This definition can be useful in some contexts, though it is unclear why raw prevalence should be a presumptively neutral baseline. Indeed, this formulation leads to the perverse outcome that spamming content may reduce amplification (through an increase in the denominator) even though it may increase distribution. For some systems a chronological baseline may make sense. This is plausible for Twitter because it was designed around a reverse-chronological feed, and several studies compare the algorithmic and chronological options on Twitter (Bandy and Diakopoulos, 2021; Huszár et al., 2021). However, there is no similarly natural baseline for systems such as YouTube, Google News, Netflix, Spotify, or Amazon where a chronological feed makes little sense. Further, purely chronological feeds can suffer from problems that make them unattractive baselines (Kantrowitz, 2021; Keller, 2021) including recency bias and the prevalence of low quality content like spam. Because of these conceptual and practical issues, "amplification" is not a well-defined measure for most recommender systems (Thorburn et al., 2022b). In the U.S., legislation targeting amplification in any of these senses is also likely to face 1st Amendment challenges (Keller, 2021).

Examples like these highlight the importance of finding new ways of bridging the knowledge divide between policy makers, specialized expertise and independent research. Possible ways forward include educational programs for policy makers and engineers alike, the embedding of technology expertise (such as TechCongress which recruits technologists and embeds them in Capitol Hill offices to inform tech-focused legislation), regulatory sandboxes which allow controlled experimentation, a strong commitment to evidence-based policy making, and other initiatives to make technical expertise more easily accessible.

## 7. Open Problems

Based on this review of values in recommender systems from a variety of perspectives, we propose the following list of open problems. Each of these is both consequential and challenging, and would benefit greatly from future work.

**Process for defining values and resolving trade-offs.** One of the overarching challenges in this area is that there is no widely-agreed process for deciding which values are prioritized and tracked, how they are measured, and how trade-offs are adjudicated. There are trade-offs or conflicts both between multiple people or stakeholders, and between different values. For example, while it is clear that users should have substantial control over the recommendations they receive, it is less clear what to do when this appears to harm oneself or others. The question of what is measured, and on which subgroups, has similarly profound implications for outcomes. In an operational sense, "values" can only come from the normative debates that take place in society (Gabriel, 2020). Methods such as multi-stakeholder metric construction (Lee et al., 2019; Stray, 2020) and user juries (Fan and Zhang, 2020) may provide a way forward, but are not well developed.

**Better measurements.** Many values are not easy to operationalize. Collections of AI-relevant metrics such as the IEEE 7010 standard (Schiff et al., 2020) provide useful compendiums, but measures developed in social science and policy may not apply directly to particular recommender contexts such as news recommendation or targeted advertising, and will need to be refined.

**Controls that people want to use.** The controls offered to users so far have been remarkably sparse and slow to develop, despite the obvious practical, policy, and ethical advantages of better control. In part this may be a problem of under-investment, but there is also a fundamental unsolved problem with recommender controls: most are never used by more than a few percent of users (Jin et al., 2017). Better control designs might improve this situation, such as immediate feedback on how changing a setting changes which items are recommended (Schnabel et al., 2020). Still, it is likely that most users will never adjust default settings, so a better strategy may be to attempt to design controls that elicit user feedback in the normal course of use, e.g. voting systems or a "respect" button (Stroud et al., 2017). It will be important to differentiate between giving users a feeling of agency and giving them effective controls (Loecherbach et al., 2021; Vaccaro et al., 2018) as both are necessary.

**Non-behavioral feedback.** At present essentially all feedback signals to recommender systems are behavioral, e.g. engagement. There are a variety of well-known problems with inferring preferences from behavior, including self-control or addiction issues, uninformed users, and the contextual meaning of choices (Anderson, 2001; Thorburn et al., 2022a). In particular, many of the values discussed above involve outcomes that are unlikely to be identifiable from on-platform behavior alone, e.g. well-being. The most promising solution is simply to ask users about their experience by repurposing survey instruments that have been developed in the social science and policy communities. Only a small fraction of users can be surveyed, but the resulting data stream can be used to build and continuously validate models that predict survey responses, which can then be used as algorithmic objectives. This is already done in industry (Goodrow, 2021; Stray, 2020) but there is essentially no public research on this emerging technique.

**Long-term outcomes.** Most recommender algorithms today are myopic in the sense of optimizing only for immediate responses. Longer-term outcomes are typically managed by product teams who monitor richer feedback and make algorithmic adjustments. While today these teams are typically optimizing for purchases, subscriptions or user retention, recommender systems could also be managed on values-relevant outcomes. However, human management will always optimize against aggregated outcomes, while algorithmic optimization can be personalized and therefore has the potential to better serve subgroups and individuals. Emerging reinforcement learning methods may make this possible (Afsar et al., 2021; Mladenov et al., 2019) but this may also require cheap and accurate individual-level proxy metrics for the outcomes of interest.

**Causal inference of human outcomes.** Recommenders may have significant effects on people and society, yet determining what those effects are remains extremely challenging due to sampling and confounding effects. Individual case studies can be instructive but are difficult to generalize to platform scale, while even large observational studies struggle to answer questions like "does social media cause depression or do depressed people use social media more?" In many cases it will not be possible to say what a platform's effect on an individual has been because there is no counterfactual, but it may be possible to measure group-level effects through on-platform experiments. The long term effects of design decisions – and the resulting outcomes for users interacting with the system – are particularly difficult to study because so many other things are happening in a user's life.

**Industry-academic research collaborations.** While a variety of external algorithmic auditing methods have been developed (Sandvig et al., 2014) many of the most important questions can only be answered in an ecologically valid setting using private platform data or on-platform experiments (Griffioen et al., 2020). Unfortunately, it is not easy for external researchers and platforms to work together due to concerns around access, privacy, security, research integrity, intellectual property, competitiveness, and cost (OECD, 2019, chap. 4; Shapiro et al., 2021; Singh, 2020a). External collaborations would benefit from the development of technical methods to enhance the privacy and security of shared data, and from legal or policy approaches that set the terms of engagement so as to protect all relevant interests.

**Interdisciplinary Policy-making.** There is a substantial gap between the policy community and the technical community (including between scholars of both). Not only is there an enormous amount of specialized knowledge unknown to each side, but they use very different terms to understand and describe the problems of recommenders. There are probably also fundamental differences in values and politics between disciplines that will have to be resolved, which is complicated by the fact that recommender policy touches fundamental rights such as privacy and freedom of expression. It is clear that effective policy-making must be collaborative and interdisciplinary, even if it is not yet clear how to achieve this.

## 8. Conclusion

Recommender systems are a profound technology that will continue to touch many aspects of individual and social life. At their best, they serve important interests of multiple parties. Consumers might want accurate information, good music, or useful products, producers might depend on recommenders to help them find their audience or customers, and platforms need to capture some of this value to continue operating. Yet recommenders can also cause a wide variety of harmful effects, or they may miss opportunities to do good. Building human values into recommender systems raises complex and consequential challenges, including philosophical, regulatory, political, sociological, and technical issues.

This paper contributes to this interdisciplinary conversation and guides further research in several ways. We have identified some values that are relevant to recommender systems, and discussed the main issues surrounding each. We have described current industrial practice including a sketch of modern recommender design, and an illustrative process for shifting a value in a production system. We discussed the challenges of measuring adherence to a value, including the difficulty of "operationalizing" or translating a value into metrics, and the types of data sources that might supply useful information on values. We feel it is too early to attempt a general theory of the value-sensitive design of recommenders, because the field is still emerging. Instead, we articulated an extensive menu of design techniques that have been applied or could likely be applied to production systems. Finally, we surveyed developing approaches to recommender regulation, identifying a substantial gap in knowledge and orientation between the technology and policy communities.

While the intersection of content personalization with individual and societal flourishing is a huge and varied area, we hope that this synthesis provides a shared language, useful starting points, and essential research directions for building human values into recommender systems.

## Appendix A: Table of Values

This list was culled from a wide variety of sources and refined through multi-stakeholder consultation as described above. It includes perspectives from various traditions and cultures, but is not intended to be a comprehensive list of the values that might be important to consider in recommender systems. Nor do we attempt to prioritize or rank these values here. Rather, this table should be viewed as one list of broadly important themes.

For each value, we provide possible interpretations in the context of recommender systems, some indicators that might be used to assess whether a system enacts or supports that value, and example designs that are relevant to that value.

| | **Value** | **Interpretation in the context of recommenders** | **Example Indicator** | **Example design** |
|---|---|---|---|---|
| 1 | Usefulness | "The purpose of recommenders is often summarized as 'help the users find relevant items'" (Jannach and Adomavicius, 2016)<br><br>A recommender should provide a useful service to the user. | Long term user retention.<br><br>User satisfaction surveys. | Show people more of what they rate highly. |
| 2 | Liberty | A platform should respect human dignity and protect human rights and freedoms (Fjeld et al., 2020, p. 61); recommenders should operationalize respect for autonomy (Varshney, 2020).<br><br>A platform should not stop users from exploring certain types of information. Users should be able to pursue their own good in their own way. | Similar mechanisms as Freedom of Expression. | Protections against arbitrary removal of content.<br><br>Eliminating threats of violence designed to compel individuals to form particular opinions, in violation of article 19 in Universal Declaration of Human Rights. |
| 3 | Freedom of Expression | "Everyone shall have the right to freedom of expression; this right shall include freedom to seek, receive and impart information and ideas of all kinds" (United Nations, 1966 art .19)<br><br>"The exercise of freedom of opinion, expression and information … is a vital factor in the strengthening of peace and international understanding" (UNESCO, 1978).<br><br>Platforms should not stop users from expressing their thoughts and opinions, or unduly suppress distribution of user posts. | "Conducting an ex-ante evaluation attempts to predict the future relationship between [human] rights and an on-going or proposed business activity." (Latonero and Agarwal, 2021) | Transparency around content which is suppressed.<br><br>Appeals processes for removed content. |
| 4 | Control | Control "includes the system's ability to allow users to revise their preferences, to customize received recommendations, and to request a new set of recommendations" (Pu and Chen, 2010, p. 18)<br><br>"Ability to determine the nature, sequence and/or consequences of technical and operational settings, behavior, specific events, and/or experiences." (IEEE, 2021, p. 17)<br><br>The platform should give users ways to control the content selected for them and it should give users ways to control how the content they create is shared. | "I feel in control of my news feed." (Vaccaro et al., 2018)<br><br>Locus of control Scale. Ex: "Other people usually control my life" (Sapp and Harrod, 1993) | Recommender controls (Harambam et al., 2019; He et al., 2016; Schnabel et al., 2020) |
| 5 | Agency, Autonomy, Efficacy | "The sense of agency can be analyzed as a compound of more basic experiences, including the experience of intentional causation, the sense of initiation and the sense of control" (Pacherie, 2007)<br><br>The platform should provide the capacity for intentional action and help users achieve their goals (Varshney, 2020)<br><br>The platform should not manipulate users for the benefit of other stakeholders. | General Self-Efficacy Scale (GSE)<br>Ex: "It is easy for me to stick to my aims and accomplish my goals" (Chen et al., 2001)<br><br>Autonomy Scale (AS): Measures components of autonomy including family loyalty autonomy, value autonomy, emotional autonomy, and behavioral autonomy (Anderson et al., 1994) | Recommender controls (Harambam et al., 2019; He et al., 2016; Schnabel et al., 2020) |
| 6 | Privacy | "The protection of personal data is of fundamental importance to a person's enjoyment of his or her right to respect for private and family life" (Council of Europe, 2020)<br><br>Users should be allowed to determine if and how their personal data is collected, processed and disseminated (Fjeld et al., 2020, p. 21; IEEE, 2021, p. 71) | GDPR Data Protection Impact Assessment (European Commission, 2016) | Controls to limit data use.<br><br>Federated recommendation algorithms (Qi et al., 2020; C. Wu et al., 2021)**.** |
| 7 | Safety, Security | The principle of 'safety' requires that an AI system be reliable and that 'the system will do what it is supposed to do without harming living beings or [its] environment.'" (Fjeld et al., 2020, p. 38)<br><br>"[System] use should not contribute to increasing stress, anxiety, or a sense of being harassed by one's digital environment." (Université de Montréal, 2018) | Sense that most people can be trusted (World Values Survey, 2020) | Hate speech moderation<br><br>Parental controls<br><br>Option to report harmful content. |
| 8 | Self-expression, authenticity | "Self-expression is a notion that is closely associated with a horde of positive concepts, such as freedom, creativity, style, courage, self-assurance, and even healing and spirituality" (Kim and Ko, 2007)<br><br>The platform should empower users to express their identity (including personality, attributes, behavior) and to decide how it is presented to others. (Kerlin, 2020, p. 19). | Correlates with (but is not identical to) quantity and quality of user content creation | Easy-to-use, attractive content creation/modification tools, e.g. SnapChat filters, TikTok tools.<br><br>Content moderation policies that allow self-expression (e.g. don't take down art because it has nudity) |
| 9 | Well-being | Well-being is a complex and multidimensional construct that encompasses many of the values we describe here and there is no universally accepted definition (Dodge et al., 2012).<br><br>Well-being may be measured by both objective and subjective factors including satisfaction with life, emotions and affect, psychological well-being, social relationships, and meaning (Linton et al., 2016)<br><br>On the platforms, users should see content that leads them to experience contentment, joy and pleasure, both ephemerally and over the long term. Platforms should help users feel satisfied with their lives. | Satisfaction with Life Scale (SWLS). Ex: “The conditions of my life are excellent” (Diener et al., 1985)<br><br>Positive Affect and Negative Affect Scales (PANAS). Ex: “Indicate the extent to which you have felt interested, distressed, excited, upset, etc over the past week” (Watson et al., 1988)<br><br>Scale of Psychological Wellbeing (SPWB). Assesses psychological functioning across domains such as autonomy, self-acceptance, purpose in life, and positive relationships (Ryff and Keyes, 1995) | Encourage active vs. passive use of social media (Verduyn et al., 2017)<br><br>Usage controls, e.g. screen time limits |
| 10 | Inspiration, Awe | “In times of uncertainty, others are sought for guidance, inspiration or motivation, to seek ideas, goals or possibilities, which can influence ambitions, choices, and achievements.” (Kerlin, 2020, p. 11)<br><br>Platforms should show users content that will inspire, motivate, or guide them. | Inspiration scale: Measures the experience of being inspired, feeling inspired and motivation to do something with that inspiration (Thrash and Elliot, 2003) | Uprank content identified by Inspiration classifier (Ignat et al., 2021) |
| 11 | Mental Health | Platforms should help users protect and improve their mental health. Platforms should not encourage unhealthy types or amounts of use.<br><br>“[A] state of well-being in which the individual realizes his or her own abilities, can cope with the normal stresses of life, can work productively and fruitfully, and is able to make a contribution to his or her community” (World Health Organization, 2004) | Warwick-Edinburgh Mental Well-Being Scale (WEMWBS) (Tennant et al., 2007)<br><br>Positive Mental Health Scale (PMH)<br>Ex: “I feel that I am actually well equipped to deal with life and it’s difficulties” (Lukat et al., 2016) | Features that help users manage their screen time, e.g. “take a break” notifications, screen time limits.<br><br>Option to report harassment or bullying. |
| 12 | Physical Health | Platforms should help users protect and improve their physical health, such as by providing accurate health information and encouraging behaviors that contribute to physical health. | WHO Wellbeing Index (WHO-5)<br>Ex: “I have felt active and vigorous” and “I woke up feeling fresh and rested” (Topp et al., 2015) | Help users develop healthy habits (exercise, diet, etc.)<br><br>Clear platform policies on removing health misinformation, ads for fake cures, etc. |
| 13 | Self-actualization, Personal Growth | Platforms should enable users to reach their full potential. | “Feeling that the things one does are worthwhile” (UK Office of National | Educational recommender systems (Dascalu et al., 2016) |
| | | "Personal growth is a continuous improvement in life in order to find purpose and meaning" (Kerlin, 2020, p. 22)<br><br>"[R]ather than have people choose the easiest option, we wish to have them develop a strong sense of determination of having selected the right path" (Knijnenburg et al., 2016) | Statistics, 2019)<br><br>European Social Survey (ESS). Ex: "To what extent [do you] learn new things in life?" (Norwegian Centre for Research Data, 2012) | |
| 14 | Recognition, Acknowledgment | Platforms should provide ways for other people to recognize a user's contributions or worth.<br><br>"Whilst some esteem needs can be met by having self-acceptance, self-worth and self-value, validation from others is also important" (Kerlin, 2020, p. 28) | Social Inclusion Scale (SIS). Ex: 'I have felt accepted by my friends'' (Wilson and Secker, 2015) | "Celebrate" reaction button (as LinkedIn has) |
| 15 | Knowledge, Informativeness | Users should see items that keep them informed about topics they care about or might care about.<br><br>"[P]ersonalised news recommendations allow the media not only to help users find relevant information, but also to inform them better and more effectively." (Helberger, 2019, p. 995)<br><br>"Curiosity drives individuals to seek stimulation, information or new experiences, serving a purpose to increase knowledge and build skills." (Kerlin, 2020, p. 18) | News knowledge quizzes (Allcott et al., 2020; Angelucci and Prat, 2021)<br><br>Curiosity and exploration inventory (CEI). Ex: "I would describe myself as someone who actively seeks as much information as I can in a new situation'' (Kashdan et al., 2004) | News recommender systems (Karimi et al., 2018)<br><br>Trending topics |
| 16 | Connection | "Individuals are driven to interact and seek social closeness … The benefits of social connections are far-reaching." (Kerlin, 2020, p. 17) | Perceived Social Support scale (PSS). Ex: "I have friends with whom I can share my joys and sorrows" and "I can talk about my problems with my friends" (Zimet et al., 1988)<br><br>Online Social Support Scale (OSSS). Ex: "Online, I belong to groups of people with similar interests" and "Online, people make me feel like I belong" (Nick et al., 2018) | Recommendations for people and groups with similar interests (Carullo et al., 2014) |
| 17 | Civic Engagement | "Working to make a difference in the civic life of one's community and developing the combination of knowledge, skills, values and motivation to make that difference." (Ehrlich, 2000, p. vi)<br><br>"Creating a public forum and optimal conditions for engagement" (Helberger, 2019, p. 1005) | Voter turnout (OECD, 2020)<br>Attendance of peaceful demonstrations (World Values Survey, 2020)<br><br>Approximate total hours a month active in voluntary organizations (World Values Survey, 2020) | Encourage people to vote (Bond et al., 2012) |
| | | | Donations to a charity in a month (National Opinion Research Center, 2013) | |
| 18 | Community, Belonging | “[T]he feeling that people matter to one another and to the group, and a shared faith that members’ needs will be met through their commitment to be together.” (McMillan and Chavis, 1986)<br><br>“The concept of members sharing mutual concern and/or love for one another” (Kerlin, 2020, p. 15) . | Sense of belonging to a neighborhood (UK Office of National Statistics, 2019)<br><br>The General Sense of Belongingness Scale (Malone et al., 2012). Ex: “I have a place at the table with others” | Recommend community-oriented groups<br><br>User affordances for donating to charitable causes. |
| 19 | Accessibility, Inclusiveness | Inclusiveness in design means that diverse people and perspectives should be involved in the design of AI systems. Inclusiveness in impact calls for a just distribution of AI benefits and harms.(Fjeld et al., 2020, pp. 51–53) | Assessing the experience of different user demographics through surveys and other methods, such as visually impaired or hard of hearing users (Berke et al., 2018; Morris et al., 2016) | Implement Web Content and Accessibility Guidelines (WCAG)<br><br>Variety of formats for cognitive impairments (e.g. screen readers, transcripts, captions) |
| 20 | Tolerance, Constructive Discourse | Tolerance “creates the opportunity for a wide range of political groups to express their ideas and to participate in public life” (Sullivan and Transue, 1999)<br><br>Polarization divides society into “us” and ”them” camps and contributes to the erosion of democracy (McCoy and Somer, 2019) | Polarization measures such as issue-position or affective polarization (Boxell et al., 2017; Iyengar et al., 2018; Yarchi et al., 2021) | Increase ideological diversity, select for civil conversations, optimize polarization measures (Stray, 2021a) |
| 21 | Duty | The notion of duty and obligation is defined in contrast to self-interest (Hursthouse and Pettigrove, 2018). | Personal control and responsibility scale (Meca, 2012) | User affordances for donating to charitable causes. |
| 22 | Care, Compassion, Empathy | “The ethic of care emphasizes the importance of context, interdependence, relationships, and responsibilities to concrete others.” (Koggel and Orme, 2010) and is “based on the recognition that self and others are interconnected” (Skoe, 2014).<br><br>These notions also appear in Confucian and Buddhist traditions (Vallor, 2016, p. 132) and in Ubuntu values where “a person is ‘a person through others’” (Ndiweni and Sibanda, 2020). | Empathy quotient (Baron-Cohen and Wheelwright, 2004). Ex: “I find it easy to put myself in somebody else’s shoes’’ and ‘’Seeing people cry doesn’t really upset me ‘’ | Best practices in suicide prevention (Robinson et al., 2016) |
| 23 | Fairness, Equality, Equity | “A morally justifiable distribution of benefits, goods, harms, and risks” (Vallor et al., 2020).<br><br>In multi stakeholder fairness, different stakeholders have different objectives and needs | Fairness metrics in classification and recommendation (Beutel et al., 2019a; Raj et al., 2020) | Add fairness measures to the recommender training or serving objectives (Beutel et al., 2019a; Mehrotra et al., |
| | | from the system (Abdollahpouri et al., 2020). | | 2018) |
| 24 | Diversity | "[D]iversity refers to the idea that in a democratic society, informed citizens construct their worldview from a diverse set of sources which helps them to make balanced and well-considered decisions." (Helberger et al., 2018, p. 192).<br><br>"'Cultural diversity' refers to the manifold ways in which the cultures of groups and societies find expression." (UNESCO, 2005 art. 4) | Slate and feed diversity metrics of various kinds (Kunaver and Požrl, 2017)<br><br>Perceived diversity and user satisfaction (Kunaver and Požrl, 2017) | Methods to increase various types of diversity, including topical and source diversity, novelty, coverage, etc. (Castells et al., 2015; Kunaver and Požrl, 2017) |
| 25 | Accountability | Accountability is "a component of the state of being responsible, alongside being answerable and being attributable" (Mattingly-Jordan et al., 2019)<br><br>AI principles documents include "verifiability and replicability," "impact assessments," "environmental responsibility," "evaluation and auditing requirements." (Fjeld et al., 2020, p. 28) | Ranking Digital Rights Corporate Accountability Index (Ranking Digital Rights, 2020) | Explanations of decisions according to clearly defined principles.<br><br>Processes for users to bring up problems and resolve them in a timely manner. |
| 26 | Transparency & Explainability | Transparency means that information provided about a system is a) meaningful, b) useful, c) accessible, d) comprehensive, and e) truthful (IEEE, 2021, p. 71)<br><br>"Transparency is instrumental to uphold intrinsic values of human autonomy and justice" (Canca, 2020, p. 20)<br><br>Desirable properties of transparency metrics (Spagnuelo et al., 2016)<br><br>"Explainability implies presenting technical information in a form understandable to, and meeting the needs of, a specific user category" (IEEE, 2021, p. 49) | System cards which provide stakeholders with an overview of different components of an ML system (Adkins et al., 2022) | Explanations for recommendations, answering questions such as "why am I seeing this?" or "Why has this been removed?"<br><br>Publicly disclose and discuss changes to moderation or ranking (Stray, 2021b). |
| 27 | Accuracy (Factuality) | Accuracy is "an AI system's ability to make correct judgements, for example to correctly classify information into the proper categories, or its ability to make correct predictions, recommendations, or decisions based on data or models" (Weiser, 2019)<br>AI systems should make "accurate information readily available" (Google, 2021) and should not spread "untrustworthy information." (Université de Montréal, 2018) | Credibility metrics including prevalence of misinformation items among recommendations, and user engagement on this material. | Misinformation classifiers (Bozarth and Budak, 2020)<br><br>Prompts to consider accuracy before sharing (Pennycook and Rand, 2021). |
| 28 | Tradition, History | "Heritage is a concept to which most people would assign a positive value. The preservation of material culture ... and intangible culture … are generally regarded as a shared common good by which everyone benefits." (Silverman and Ruggles, 2007) | Percentage of locally produced content. | Promotion of cultural events (Chaopreecha, 2019) |
| | | Practices, symbols, ideas, and beliefs that are developed by groups and represent their shared experience and fate. They often take the form of religious rites, beliefs, and norms of behavior (Schwartz, 2012, p. 6).<br><br>Rituals or conventional norms are one of the five central virtues of Confucianism (Csikszentmihalyi, 2020) | | |
| 29 | Environmental Sustainability | "Respect for environment and natural habitat, efficiency, maintainability, operability, supportability, reliability, durability, resilience, forgiveness, robustness, redundancy, reusability, reconfigurability, simplicity, economy, renewability" (IEEE, 2021, p. 70)<br><br>"Develop and scale up carefully assessed technologies, infrastructure and actions that reduce climate change and its associated risks." (UNESCO, 2017) | Product level carbon footprints (Meinrenken et al., 2012) | Recommender systems can potentially direct users toward climate-friendly options (Rolnick et al., 2019, p. 25)<br><br>Reporting on the energy used by ML models (Henderson et al., 2020) |
| 30 | Progress | The International Covenant on Civil and Political rights includes the right of people to "freely pursue their economic, social and cultural development" (United Nations, 1966) as expanded in the Declaration on Social Progress and Development (United Nations, 1969). | UN Human Development Index (Stanton, 2007) | To the best of our knowledge, none |
| 31 | Labor | Meaningful work is both significant and positive in valence (Rosso et al., 2010). Job insecurity, low employability, and unemployment are all detrimental to well-being, beyond the effects of income loss (Green, 2011) | Work and Meaning Inventory (Steger et al., 2012) | Ensure users see recommendations for jobs that they would want. |

## Acknowledgements

Thank you to Chloe Bakalar, Daphne Keller, Joelle Pineau, and Andrea Wong for helpful feedback.